\documentclass[fleqn,10pt]{wlscirep}

\newcommand{\vp}{\varphi}
\newcommand{\w}{\omega}

\newcommand{\e}{\varepsilon}

\newcommand{\dd}{\mathrm{d}}

\newcommand{\beginsupplement}{
	\setcounter{table}{0}
	\renewcommand{\thetable}{S\arabic{table}}
	\setcounter{figure}{0}
	\renewcommand{\thefigure}{S\arabic{figure}}
}

\title{Inferring the phase response curve from observation of a continuously perturbed oscillator}

\author[1,2]{Rok Cestnik}
\author[1,3,*]{Michael Rosenblum}
\affil[1]{Department of Physics and Astronomy, University of Potsdam, 
Karl-Liebknecht-Str. 24/25, D-14476 Potsdam-Golm, Germany}
\affil[2]{Department of Human Movement Sciences, MOVE Research Institute
Amsterdam, Vrije Universiteit Amsterdam, van der Boechorststraat 9,
Amsterdam, Netherlands}
\affil[3]{Department of Control Theory, Nizhny Novgorod State University,
Gagarin Av. 23, 606950, Nizhny Novgorod, Russia}

\begin{abstract}
Phase response curves are important for analysis and modeling of 
oscillatory dynamics in various applications, particularly in neuroscience. 
Standard experimental technique for determining them requires 
isolation of the system and application of a specifically designed input. 
However, isolation is not always feasible and we are compelled to observe the 
system in its natural environment under free-running conditions. 
To that end we propose an approach relying only on passive observations of the system and its input. 
We illustrate it with simulation results of an oscillator driven by a stochastic force.
\end{abstract}

\begin{document}

\flushbottom
\maketitle
\thispagestyle{empty}

\section*{Introduction}
Phase response curve (PRC), also known as phase resetting or phase sensitivity 
curve, is a basic characteristic of a limit cycle 
oscillator~\cite{Winfree-80,Glass-Mackey-88,Rinzel-Ermentrout-98,PRC-Scholarpedia}. 
This curve describes 
variation of the phase $\vp$ of the system in response to a 
weak external perturbation $p(t)$:
\begin{equation}
	\dot{\vp} = \w + Z(\vp)p(t) \;.
	\label{eq1}
\end{equation}
Here $\w$ is the natural frequency and $Z(\vp)$ is the oscillator's PRC. 
Thus, knowledge of $Z(\vp)$ and $\w$ completely determines the phase dynamics 
for any given weak perturbation.  
For example, it allows one to determine -- 
at least numerically -- 
the domain of locking to an external force of a given amplitude.
Hence, techniques for efficient experimental determination of PRC 
are in high demand.

If it is possible to isolate the oscillator and apply a specially designed 
perturbation, then determination of PRC is straightforward. 
To this goal, the experimentalist has to "kick" the system with 
short and weak pulses at different values of the oscillator's phase and look for 
subsequent variation of one or several oscillation periods, i.e., for the evoked asymptotic phase 
shift~\cite{PRC-Scholarpedia, Glass-Mackey-88, Izhikevich-2007, Guevara-1986, Imai-2017}.
If the oscillator is noisy -- and all real-world systems are -- then 
stimulation for each $\vp$ shall be performed many times and the results shall 
be averaged over the trials.
Next, the whole procedure shall be repeated for different amplitudes of the 
stimulation in order to check whether the obtained PRC linearly scales with 
the amplitude: this would indicate that the chosen stimulation is sufficiently weak
(we remind that in theory the PRC is defined for infinitesimally weak perturbations~\cite{Rinzel-Ermentrout-98, Winfree-80}).  

If one has no control over the applied input and instead has to rely on passive observations, 
the problem of invoking the PRC is in general not solved and the method will depend on the input. 
Possibly the best-case scenario is when the input is weak, spiky and with a frequency 
several times lower than that of the oscillator. 
Then, as long as the oscillator does not synchronize with the input, the spikes arrive at different phases and never more than one per period. 
This results in a focused perturbation which provides an estimate of phase shifts 
for different oscillator 
phases, like in the standard technique \cite{PRC-Scholarpedia, Glass-Mackey-88, Izhikevich-2007, Guevara-1986, Imai-2017}. 
If, however, the frequency of the input spike train is higher than that of the oscillator then 
each period on average gets perturbed more than once, which considerably adds to the 
complexity of the problem~\cite{Cestnik-Rosenblum-17}. 
In general the input may be arbitrary, in particular, it can be a continuous noise-like signal. 
If both such input and output of the system are measured, then one can infer the PRC using
the idea of Spike-Triggered Average (STA)~\cite{Galan-PRL_2007}. 
As has been demonstrated theoretically and numerically for the Hodgkin-Huxley neuronal model,
for weak delta-correlated input the STA is proportional to the derivative of the PRC~\cite{Galan-PRL_2007}.
Recent more practical algorithms exploit weighted STA (WSTA): they imply rescaling of the input within each 
inter-spike interval to the same length and subsequent averaging with the weights, determined
by the length of these intervals~\cite{Ota-PRL-2009, Imai-2017}. Another approach is to use STA
with a specific optimal colored noise input~\cite{Morinaga-2015}.

In this paper we introduce a method for inferring the PRC by fitting the phase model (\ref{eq1})
to the observed time series. This nonlinear problem is solved by an iterative procedure 
that provides frequency $\w$ and PRC $Z(\vp)$ as well as instantaneous phase $\vp(t)$. 
An important novel feature of our algorithm is a built-in error estimation
that allows us to monitor the goodness of the reconstruction.
We demonstrate the efficiency of the proposed method on several oscillator models driven by 
correlated noise and compare it with the techniques from Refs.~\cite{Ota-PRL-2009, Imai-2017}. 
We show that our technique outperforms them in case of very short time series, not very short
correlation and not very weak amplitude of the input.

\section*{Results}
\subsection*{Technique}

Suppose we perturb the oscillator under study with
stimulation $p(t)$ and record its output $x(t)$. 
For example, $x(t)$ corresponds to the membrane potential of a cell.
Since we assume that the oscillator has a stable limit cycle, without 
perturbation the process $x(t)$ would be periodic. In practice, because noise 
is inevitable, the oscillation would be almost periodic. 
The technique of extracting the PRC crucially depends on the properties of the signal $x(t)$ as well as those of the perturbation $p(t)$. 
We first discuss the relatively simple case 
where $x(t)$ is a smooth signal, suitable for phase estimation with the conventional technique of 
embedding the signal via the Hilbert 
transform~\cite{Gabor-Hilbert, Hilbert-Scholarpedia, Pikovsky-Rosenblum-Kurths-01} 
and performing the protophase-to-phase transformation~\cite{Kralemann_et_al-08}. 

Once an estimate of the phase $\vp(t)$ is obtained, the instantaneous frequency $\dot\vp(t)$ 
can be numerically estimated, 
and then, representing the PRC as a finite Fourier series,
\begin{equation}
Z(\varphi) = a_0 + \sum\limits_{n = 1}^{N} \left [ a_n \cos(n \vp) + b_n \sin(n \vp)\right ] \;,
\label{eq2}
\end{equation} 
we obtain from Eq.~(\ref{eq1}) a system of linear equations with 
respect to the unknown frequency $\w$ and the Fourier coefficients $a_n$, $b_n$.
Provided we have a good resolution and long enough time series, 
this system is overdetermined and can be solved by means of optimization, e.g., 
least squares minimization. 
Notice that this can be done for an arbitrary form of perturbation $p(t)$. 
Thus, a common approach to the problem revolves around estimating the phase $\varphi$ 
and its derivative $\dot{\varphi}$ (the latter requires numerical differentiation and is 
therefore sensitive to noise). 
For cases when the signal has a complex form and there is no unique center of rotation in 
the Hilbert representation, alternative ways of estimating $\varphi$ 
and $\dot{\varphi}$ shall be used. 
For example, for neural oscillations the phase is often obtained via simple linear 
interpolation between spikes~\cite{Galan-Ermentrout-Urban-05, Ota-PRL-2009, Imai-2017}. 
In a different setup, like in studies of the interaction between the respiratory and cardiac oscillators 
several markers within each cycle of the electrocardiogram are used to determine 
the phase~\cite{Kralemann_et_al-13}. 
Such \textit{ad hoc} techniques proved being useful, albeit they tend to be setup-specific, 
lack generality, and provide only zero-order approximation of the phase.

Here we introduce a technique that does not require computation
of continuous phase $\vp$ and instantaneous frequency $\dot{\varphi}$ from the data.
For our approach, in addition to continuous observation of the input forcing $p(t)$, 
it is sufficient to define one well-defined event within each cycle of $x(t)$ such that 
the times $t_m$ of these events can be considered as the instants of return to a 
Poincar{\'e} surface of section. In the simplest case this event can correspond to a condition $x(t) = \text{const}$, 
while generally the problem of choosing a proper Poincar{\'e} section is not so trivial and is discussed in more detail below.
Correspondingly, these events can be assigned the same 
value of the phase which without loss of generality can be set to zero, $\varphi(t_m) = 0$.
For these events one can choose spikes, threshold crossings, or even bursts (an example can be found in Supplementary material, Fig.~\ref{sup:burst}).
So, in analysis of spiking neurons it is natural to use spikes since they can be reliably 
detected while the signal between them can be dominated by noise. 
The phase increase within each inter-event interval, $T_m = t_{m+1}-t_m$, 
is defined to be $2\pi$. 
 
In what is the \emph{first key step} of our approach, 
we integrate Eq.~(\ref{eq1}) over each interval $T_m$, replacing $Z(\varphi)$ 
by its Fourier representation via Eq.~(\ref{eq2}), 
\begin{equation}
    2\pi = \w T_m + a_0\int\limits_{t_m}^{t_m+T_m}p(t)\dd t+
	\sum\limits_{n = 1}^{N} \left [ 
	a_n\int\limits_{t_m}^{t_m+T_m} p(t) \cos[n\varphi(t)] \dd t
	+b_n\int\limits_{t_m}^{t_m+T_m} p(t) \sin[n\varphi(t)] \dd t
	\right ] \;.
	\label{eq4}
\end{equation}
Certainly, the computation of the right hand side requires knowledge of the phase.
Here we point out that in the limit of vanishing perturbations, the phase between events
grows nearly linearly. 
Hence, for a sufficiently weak $p(t)$ the problem can be quite precisely
solved by linearly interpolating the phase $\varphi(t) = 2\pi (t-t_m)/T_m$, 
inserting it in Eq.~(\ref{eq4}), numerically computing all the integrals, and solving the 
linear system by means of optimization, cf.~\cite{Galan-Ermentrout-Urban-05}.
However, for stronger driving amplitudes the increased inaccuracy of the linear approximation
unavoidably translates to inaccuracy of the solution. 
We address this with the \emph{second key step} in our technique, where
the phase is estimated through an iterative procedure, by solving 
systems (\ref{eq1}) and (\ref{eq4}) in turns. 
For the zeroth approximation of the phase we linearly interpolate it between events,
 $\varphi^{(0)}(t) = 2\pi (t-t_m)/T_m$. 
Then, inserting it in Eq.~(\ref{eq4}) and solving the system we obtain the 
first approximation of the solution $\w^{(1)}, Z^{(1)}$. 
The next approximation of the phase is obtained by numerically integrating Eq.~(\ref{eq1}), 
i.e., by solving $\dot\vp^{(1)}=\w^{(1)}+Z^{(1)}[\vp^{(0)}(t)]p(t)$ separately for each 
inter-event interval, taking $\vp^{(1)}(t_m) = 0$ for initial conditions. 
Since the phase $\varphi^{(0)}$, natural frequency $\w^{(1)}$, and PRC $Z^{(1)}$ 
are known only approximately, 
the computed phase at the end of the interval, $\vp^{(1)}(t_m+T_m)=\psi_m^{(1)}$, 
generally differs from the correct value $2\pi$.  
Therefore, we \textit{rescale} the phase within each interval: 
$\vp^{(1)}(t)\rightarrow 2\pi\vp^{(1)}(t)/\psi_m^{(1)}$. 
This rescaled phase is then used to obtain the second approximation of the solution 
$\w^{(2)}$, $Z^{(2)}$, 
and then the whole procedure can be repeated to obtain further ones. 
As illustrated below, the iterations typically converge to the true solution and 
the quantities $\psi_m^{(k)}$ 
can be used to monitor the convergence. 

\subsection*{Numerical tests}
First we mention that for clarity of presentation we always set the natural 
frequency of the oscillator $\w$ to $2\pi$, thus ensuring the period of an unperturbed oscillator is $T = 1$  
(with oscillators where natural frequency is not an explicit parameter this is done by rescaling time). 
Also, throughout this paper, unless specified differently, we use $k=10$ iteration steps, $N = 10$ Fourier harmonics, time of simulation $t_\text{sim} = 500$ and time step $\dd t = 0.001$. 

\subsubsection*{Phase model: proof of concept}
We first test the technique on models, where both the true phase and the PRC are known. 
For this purpose we simulate Eq.~(\ref{eq1}) with PRCs given as 
\begin{equation}
Z(\vp)=(1-\cos(\vp)) \exp{\left(3[\cos(\vp-\pi/3)-1]\right )}\;,
\label{prc1}
\end{equation}
or
\begin{equation}
Z(\vp)=-\sin(\vp) \exp{\left(3[\cos(\vp-0.9\pi)-1]\right )}\;.
\label{prc2}
\end{equation}
These curves model PRC of type I and II respectively, as they are classified in the 
context of neuronal modeling~\cite{Hansel-95,Cestnik-Rosenblum-17}. 
Type I curves are non-negative which means that every stimulus from an excitatory connection shortens the period duration, 
while type II curves are positive at some phases and negative at others meaning that a stimulus may shorten or lengthen the period depending on when it arrives. 
We take the input to the system to be an Ornstein–Uhlenbeck process~\cite{ornstein-uhlenbeck}
\begin{equation}
\dot{p} = -\frac{p}{\tau} + \e \sqrt{2/\tau}\ \xi(t)\;,
\label{driving}
\end{equation}
where $\xi$ is Gaussian white noise $\langle\xi(t)\xi(t')\rangle=\delta(t-t')$, 
while $\e$ and $\tau$ are the amplitude and correlation time of the driving signal, $\langle p(t) p(t')\rangle=\e^2\ e^{-\frac{|t-t'|}{\tau}}$. 
At this point we mention that for the same value of $\e$ different oscillators are perturbed by different amounts 
because they have different PRCs and frequencies. From Eq.~(\ref{eq1}) we see that the effect of perturbation 
is proportional to $\e \lVert Z \rVert / \w$ where 
$\lVert \cdot \rVert$ stands for norm (we use the $L_2$ functional norm, see Methods). 
Therefore, to compare the results for different test models and since all frequencies are set to $2\pi$, we specify driving strength with the quantity $\e \lVert Z \rVert$.  

Spike data is generated using the condition $\vp(t_m) = m\cdot 2\pi$.
Then, we perform $k$ iterations of the reconstruction procedure 
and compare the results against the data. 
To quantify the quality of the reconstruction we compute the difference 
between the recovered and true PRCs, $\Delta_Z$, and average deviation of $\psi_m$ from $2\pi$, $\Delta_\psi$ 
(see  Eqs.~(\ref{err_prc}) and (\ref{err_psi}) in Methods). 
$\Delta_Z$ should be compared to 1, where $\Delta_Z \ll 1$ means the reconstructed PRC matches the true one closely, 
while $\Delta_Z \approx 1$ signals that the reconstructed PRC does not resemble the true one at all. 
$\Delta_\psi$ is to be compared to a related measure of event irregularity, $\Delta_{\psi_T}$ 
(see Eq.~(\ref{err_psi0}) in Methods), 
where $\Delta_\psi \ll \Delta_{\psi_T}$ means that our reconstruction can reproduce the data well, 
while $\Delta_\psi \approx \Delta_{\psi_T}$ signals that, statistically speaking, the reconstructed PRC 
and frequency are not able to reproduce the data any more accurately 
than a perfectly periodic oscillator with frequency $\langle \w \rangle = \langle \frac{2\pi}{T_m} \rangle$.

\begin{figure}[!tbh]
\centering
\includegraphics[width=0.85\columnwidth]{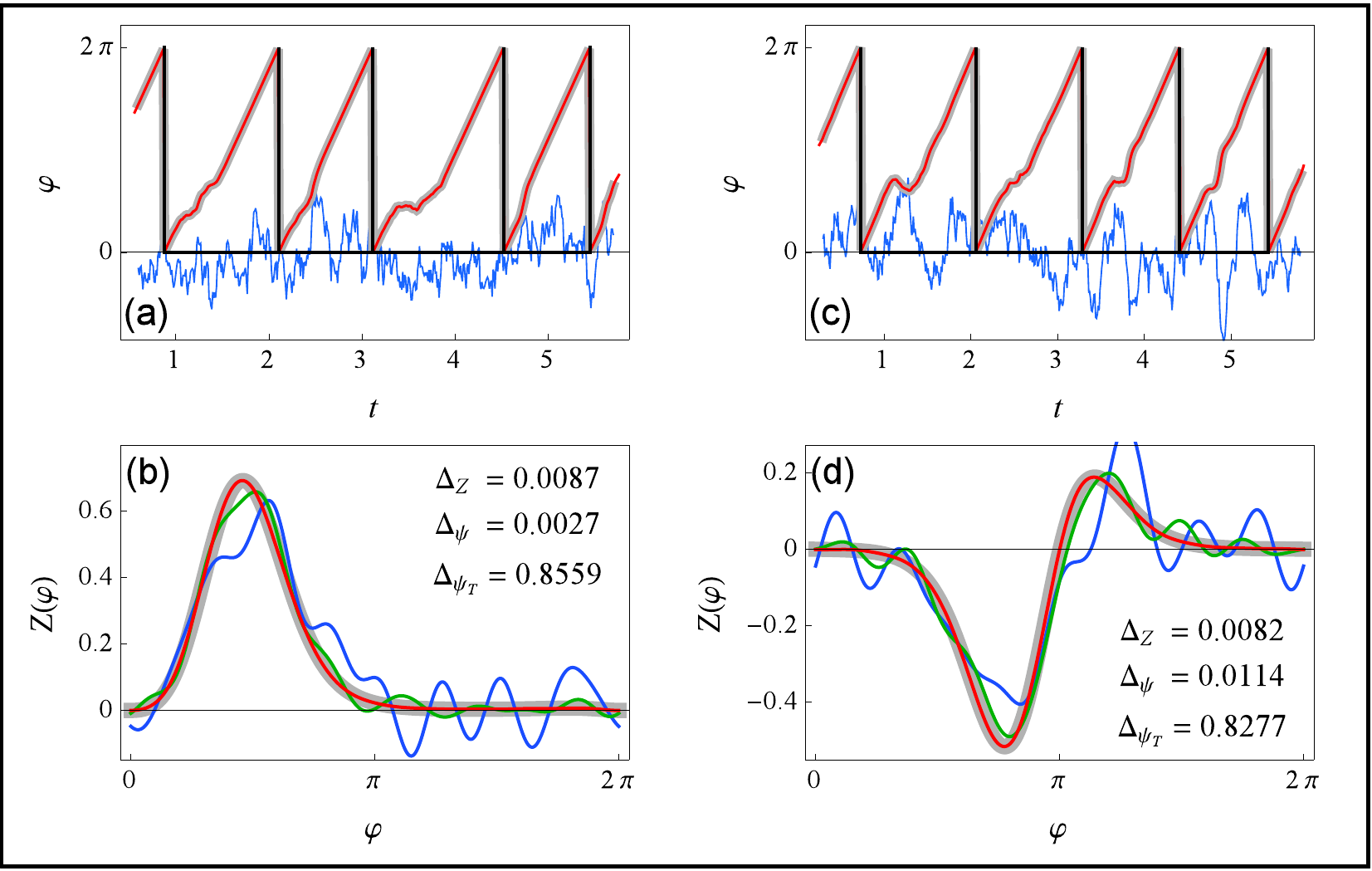}
\caption{A reconstruction of phase $\varphi$ and PRC $Z$ for PRC type I (a, b) and type II (c, d). 
In the top plots (a, c), true phase is plotted in thick gray, reconstructed phase in red, driving signal in blue (scaled for being visually comparable to the phase) and times of spikes $t_m$ in black. 
In the bottom plots (b, d), true PRC is plotted in thick gray, first iteration $Z^{(1)}$ in blue, second $Z^{(2)}$ in green and tenth $Z^{(10)}$ in red. Parameters are $\e \lVert Z \rVert = 5$ and $\tau = 0.1$.}
\label{fig:1}
\end{figure}

\begin{figure}[!tbh]
\centering
\includegraphics[width=0.85\columnwidth]{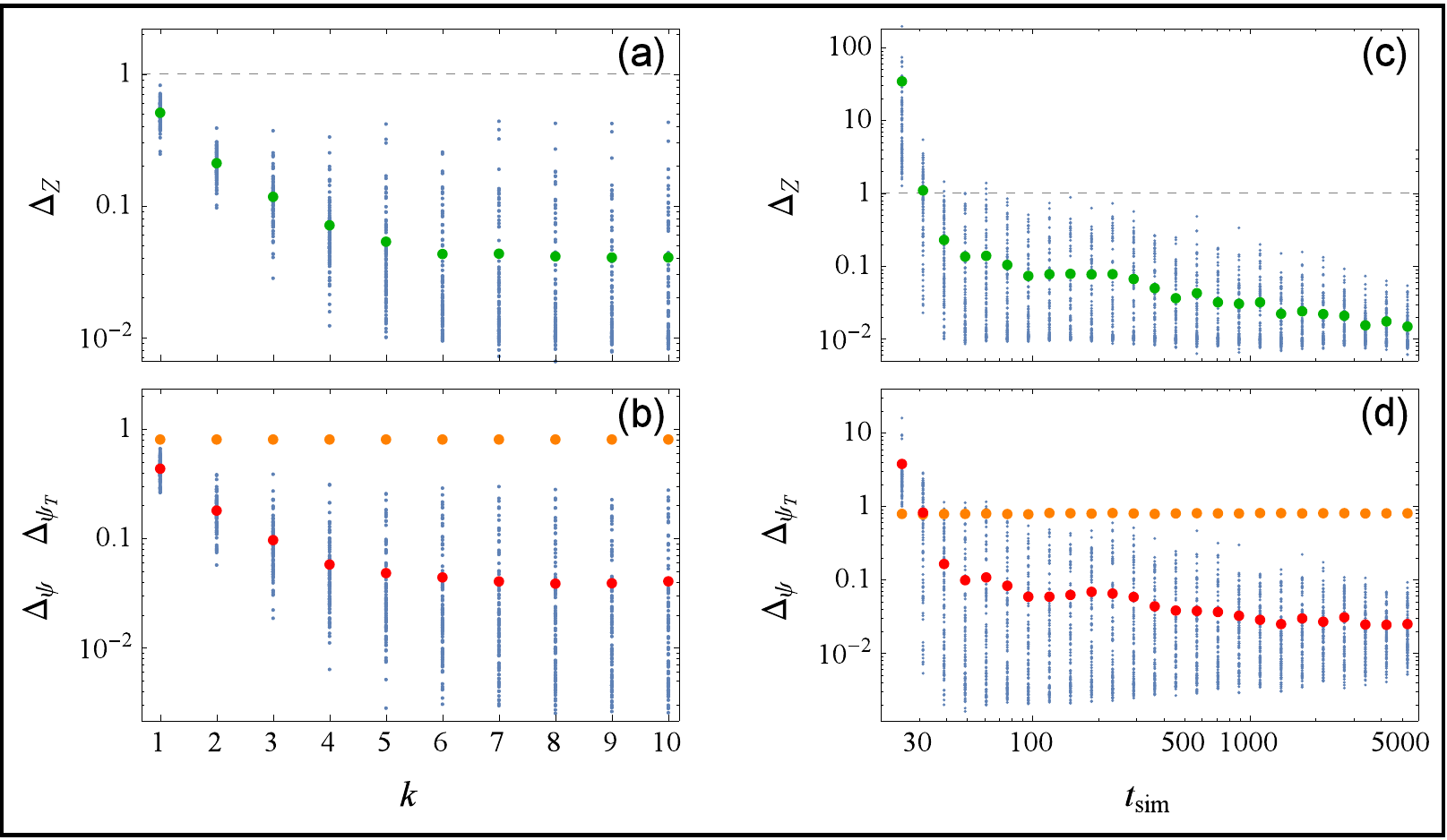}
\caption{Errors of reconstruction $\Delta_Z$ and $\Delta_{\psi}$, Eqs.~(\ref{err_prc}) and (\ref{err_psi}), 
versus iteration number $k$ (a, b) and versus the duration of data recordings $t_\text{sim}$ (c, d), 
for PRC type II. 
For each $k$ and each $t_\text{sim}$, 100 points corresponding to different realizations of noise are 
plotted with blue dots and their average in colors: $\Delta_Z$ in green, $\Delta_{\psi}$ in red 
(and $\Delta_{\psi_T}$ in orange for reference). 
Parameters are $\e \lVert Z \rVert = 5$, $\tau = 0.1$, $t_\text{sim} = 500$ (a, b) and $k = 10$ (c, d).}
\label{fig:2}
\end{figure}

In Fig.~\ref{fig:1} we show a reconstruction example for the first test. 
We use fairly strong driving $\e \lVert Z \rVert = 5$ and moderate correlation $\tau = 0.1$ so that the phase alterations are visible in the plot. 
Notice that the error measures indicate a good reconstruction, i.e., 
$\Delta Z \ll 1$ and $\Delta_\psi \ll \Delta_{\psi_T}$. 
Notice also that the initial PRC estimate $Z^{(1)}$ is quite imprecise while further iterations 
improve in precision and converge to the true curve. 
To see this convergence in more detail we plot the error measures for each iteration $k$ in 
Fig.~\ref{fig:2} (a, b), for PRC type II. 
We also investigated how the error depends on the duration of data recordings $t_\text{sim}$ 
and found that a few hundred periods suffice for a reliable reconstruction, 
see Fig.~\ref{fig:2} (c, d).

\subsubsection*{The Morris-Lecar neuron: effect of amplitude $\e$ and correlation time $\tau$}
Our next test model is the Morris-Lecar neuronal 
oscillator~\cite{Morris-Lecar-81,Rinzel-Ermentrout-98} 
\begin{equation}
\begin{split}
\dot V =& \ I -g_L~(V-V_L) -g_K~w~(V-V_K)    
       -g_{Ca}~m_{\infty}(V)~(V-V_{Ca}) + p(t) \;, \\[1ex]
\dot w =& \ \lambda(V)~(w_{\infty}(V)-w)\;,
\end{split}
\label{ml}
\end{equation}
where 
\begin{equation}
\begin{split}
m_{\infty}(V)&=[1+\tanh{((V-V_1)/V_2)}]/2\;, \\
w_{\infty}(V)&=[1+\tanh{((V-V_3)/V_4)}]/2\;, \\
\lambda(V)&=\cosh{[(V-V_3)/(2V_4)]}/3\;,
\end{split}
\label{ml2}
\end{equation}
and the parameters are:
$I=0.07$,
$g_L=0.5$, $g_K=2$, $g_{Ca}=1.33$, $V_1=-0.01$, $V_2=0.15$, $V_3=0.1$, $V_4=0.145$, $V_L=-0.5$, $V_K=-0.7$ and $V_{Ca}=1$. 
A reconstruction depiction similar to Fig.~\ref{fig:1} can be found in Supplementary material, Fig.~\ref{sup:all}.

\begin{figure}[!tbh]
\centering
\includegraphics[width=1.0\columnwidth]{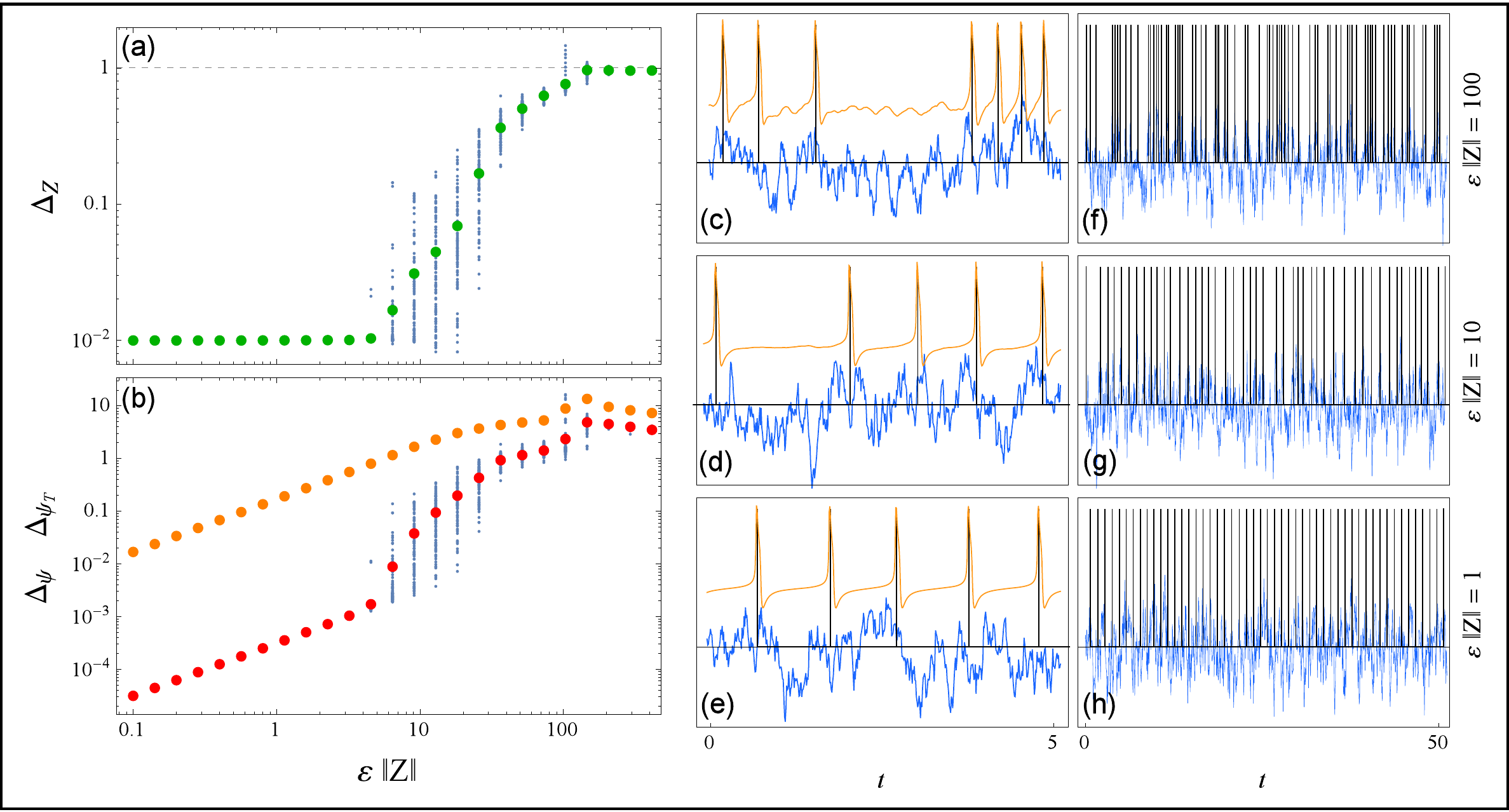}
\caption{The dependence of errors of reconstruction $\Delta_{Z}$ and $\Delta_{\psi}$, Eqs.~(\ref{err_prc}) and (\ref{err_psi}), on driving amplitude $\e$ for the Morris-Lecar neuronal oscillator Eq.~(\ref{ml}). 
In (a, b) for each value of $\e$, 100 points corresponding to different realizations of noise are plotted with blue dots and their average in colors: $\Delta_Z$ in green, $\Delta_{\psi}$ in red (and $\Delta_{\psi_T}$ in orange for reference). 
Shown in (c -- h) are segments of signal on two time scales for three amplitudes of driving $\e \lVert Z \rVert$. In orange the voltage $V(t)$ (only in (c -- e)) and in blue the input signal $p(t)$, scaled for being visually comparable. 
The times of zero phase $t_m$ are marked with vertical black lines. The correlation time is $\tau = 0.1$.}
\label{fig:3}
\end{figure}

We investigate how amplitude of driving $\e$ effects the reconstruction. 
For that purpose we simulate a forced oscillator with different values of $\e$ 
and compute the errors of reconstruction $\Delta_Z$, Eq.~(\ref{err_prc}), and $\Delta_\psi$, Eq.~(\ref{err_psi}), see Fig.~\ref{fig:3}. 
The true PRC of the Morris-Lecar system is obtained via the standard technique, see Methods and 
Ref.~\cite{Imai-2017}.
For weak driving the PRC error $\Delta_Z$ is small and independent of amplitude, while 
$\Delta_\psi$ and $\Delta_{\psi_T}$ scale approximately linearly with amplitude and maintain a constant ratio, 
$\Delta_\psi/\Delta_{\psi_T} \approx \text{const} \ll 1$. The reconstruction works well, as expected. 
As the driving amplitude increases, at some point $\Delta_Z$ begins to grow and $\Delta_\psi$ starts to 
approach $\Delta_{\psi_T}$, 
a sign of declining reconstruction quality (in Fig.~\ref{fig:3} this happens around $\e \lVert Z \rVert \approx 5$). 
In this regime the distribution of errors (both $\Delta_Z$ and $\Delta_\psi$) widens, 
meaning that reconstruction quality for the same parameters may vary considerably from trial to trial. 
As the driving gets very strong (around $\e \lVert Z \rVert \approx 100$ in Fig.~\ref{fig:3}) the reconstruction does not work, i.e., 
$\Delta_Z \approx 1$ and $\Delta_\psi \approx \Delta_{\psi_T}$. 
As a side note, in Fig.~\ref{fig:3} (b) one can observe what looks like a change in behavior 
around $\e \lVert Z \rVert \approx 100$, 
where there is a jump in $\Delta_\psi$ and $\Delta_{\psi_T}$. 
This is due to the test model we use, Morris-Lecar, and its behavior when strongly stimulated (at that amplitude the system spends some time in the vicinity of the fixed point) 
and is not a consequence of our technique (in Supplementary material one can see analogous plots for other oscillators where this is not observed, Fig.~\ref{sup:all}). 

\begin{figure}[!tbh]
\centering
\includegraphics[width=1.0\columnwidth]{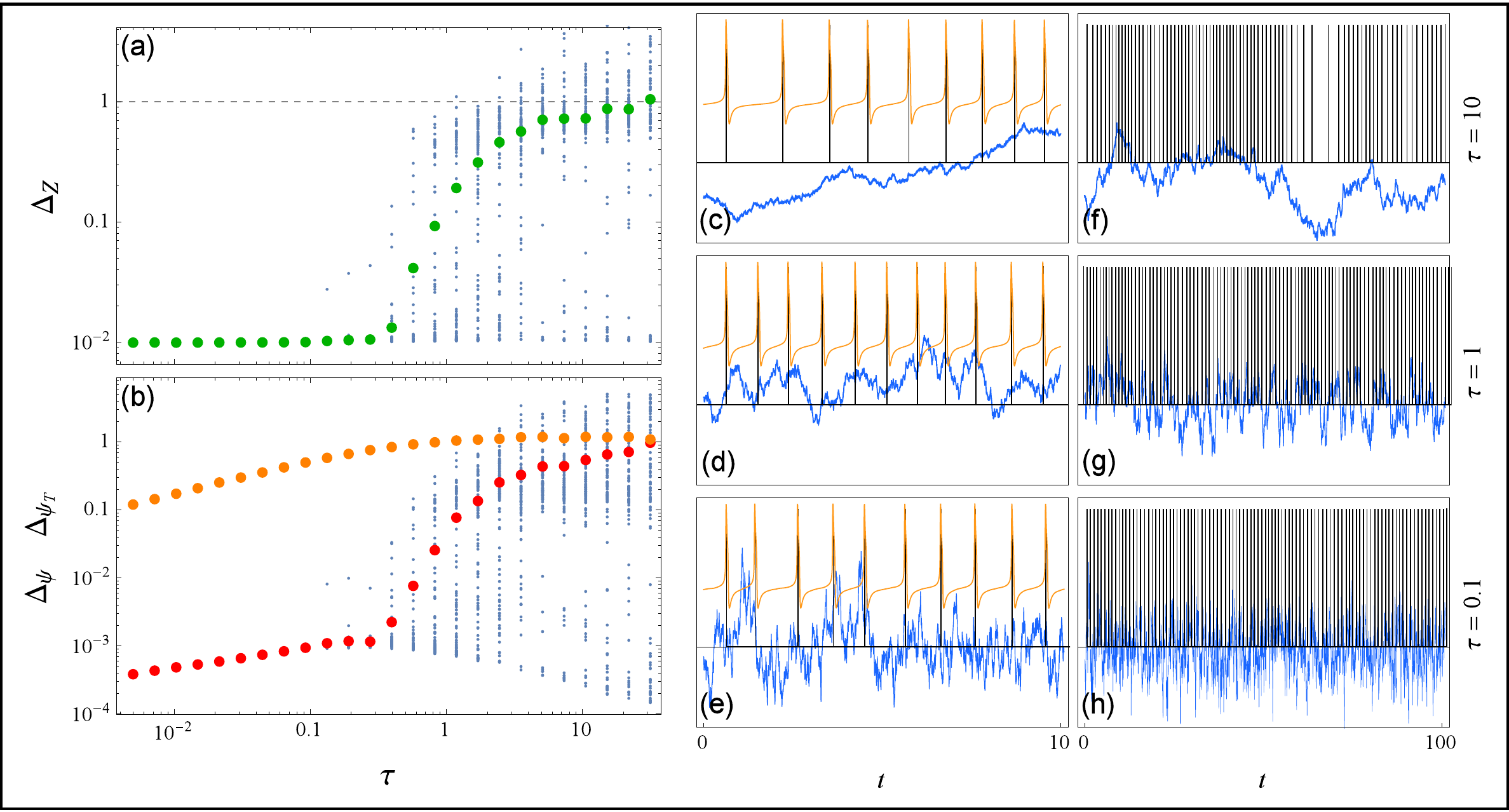}
\caption{The dependence of errors of reconstruction $\Delta_{Z}$ and $\Delta_{\psi}$, Eqs.~(\ref{err_prc}) and (\ref{err_psi}), on correlation time $\tau$ of the driving for the Morris-Lecar oscillator, Eq.~(\ref{ml}). 
In (a, b) for each value of $\tau$, 100 points corresponding to different realizations of noise are plotted with blue dots and their average in colors: $\Delta_Z$ in green, $\Delta_{\psi}$ in red (and $\Delta_{\psi_T}$ in orange for reference). 
Shown in (c -- h) are segments of signal on two time scales for three values of correlation time $\tau$. In orange the voltage $V(t)$ (only in (c -- e)) and in blue the input signal $p(t)$, scaled for being visually comparable. 
The times of zero phase $t_m$ are marked with vertical black lines. The driving strength is $\e \lVert Z \rVert = 3$.}
\label{fig:4}
\end{figure}

In the same way as for the amplitude, we now investigate how correlation time $\tau$ effects the reconstruction, see Fig.~\ref{fig:4}. 
The reconstruction works best for short correlation times: $\Delta_Z$ as well as the ratio $\Delta_\psi/\Delta_{\psi_T}$ are small and roughly constant, 
but in this case $\Delta_\psi$ and $\Delta_{\psi_T}$ grow sublinearly with $\e$. 
Similarly to what we have seen in the previous figure, as we increase $\tau$ there comes a point where 
$\Delta_Z$ starts to grow and $\Delta_\psi$ starts to approach $\Delta_{\psi_T}$ (this happens around $\tau \approx 0.4$ in Fig.~\ref{fig:4}). 
Likewise the distribution from this point on widens, indicating that reconstruction quality varies from trial to trial. 
As we continue to increase $\tau$ there comes a point at which average $\Delta_Z$ reaches 1 and $\Delta_\psi$ reaches $\Delta_{\psi_T}$ 
(this happens around $\tau \approx 30$ in Fig.~\ref{fig:4}). 
Interestingly, we can see that while the error averages in general reflect a bad reconstruction, there persists a branch of reconstructions that maintains a low error 
which is even slightly decreasing with $\tau$. 
This means that with a small probability we can still get a good reconstruction from slow varying input 
(an example can be found in Supplementary material, Fig.~\ref{sup:slow}), 
and by monitoring the solution errors we can distinguish successful reconstructions from the unsuccessful ones. 
A similar qualitative description is valid for other oscillators in this context as well 
(analogous analysis on other oscillators can be found in Supplementary material, Fig.~\ref{sup:all}). 
Here it is worth mentioning that while our chosen input, Eq.~(\ref{driving}), yields white noise with zero 
intensity in the limit $\tau \rightarrow 0$, we have tested our procedure for white noise with finite intensity 
and it works as well (an example can be found in Supplementary material, Fig.~\ref{sup:white}).

Up to this moment we have been postponing a discussion of a crucial problem, namely how to determine
the instants $t_m$ corresponding to zero phase, $\vp(t_m)=0$. 
For spiky data where only the times of events can be reliably detected (so that the data is represented
by a point process) there is no other option but to use the times of events.
This corresponds to pure phase models, where this problem does not
exist and the values $t_m$ are obtained in the course of simulation. 
This is not the case for full models, where both phases and amplitudes shall be taken into account.
So, for the analysis of the Morris-Lecar model we used a simple threshold rule and defined zero phase 
from the condition $x(t_m)= x_\text{min} + 0.9(x_\text{max}-x_\text{min})$, 
$\dot{x}(t_m) < 0$, where $x_\text{min}$ and $x_\text{max}$ are the minimal and maximal measured 
value of the observed signal.
However, this is not exact: in fact, the proper section of the limit cycle shall coincide with a
line (or, generally, surface) of constant phase, called isochrone~\cite{Guckenheimer-75}, but in 
experimental conditions the model equations and isochrones are not known \textit{a priori}.
Nevertheless, for neuronal oscillators this is not a complication if the threshold is chosen
close to the maximum of the pulse, i.e., in the domain of fast motion along the limit cycle.
Here the density of isochrones is low, i.e., phase gradient in the direction perpendicular to the isochrones 
is small (for an illustration of the isochrone structure see Supplementary material, Fig.~\ref{sup:isos}b).
Respectively, the error due to deviation of the line $x=\text{const}$ from the isochrone is small as well.
For general oscillators this is not true, and the problem requires a special treatment,
illustrated by the following example. 

\subsubsection*{The van der Pol oscillator: importance of choosing a proper Poincar{\'e} section}
Our third test system is the van der Pol oscillator~\cite{van-der-Pol-26} 
\begin{equation}
\ddot{x} - 2(1-x^2)\dot{x} + x = p(t)\;.
\label{vdp}
\end{equation}

With this model we address the importance of choosing an appropriate Poincar{\'e} section 
for determination of the beginnings of each cycle, i.e., instants $t_m$ corresponding to zero 
phase $\vp(t_m)=0$. Here we make use of the ability  of our technique to monitor 
the error of reconstruction.
 
\begin{figure}[!tbh]
\centering
\includegraphics[width=1.0\columnwidth]{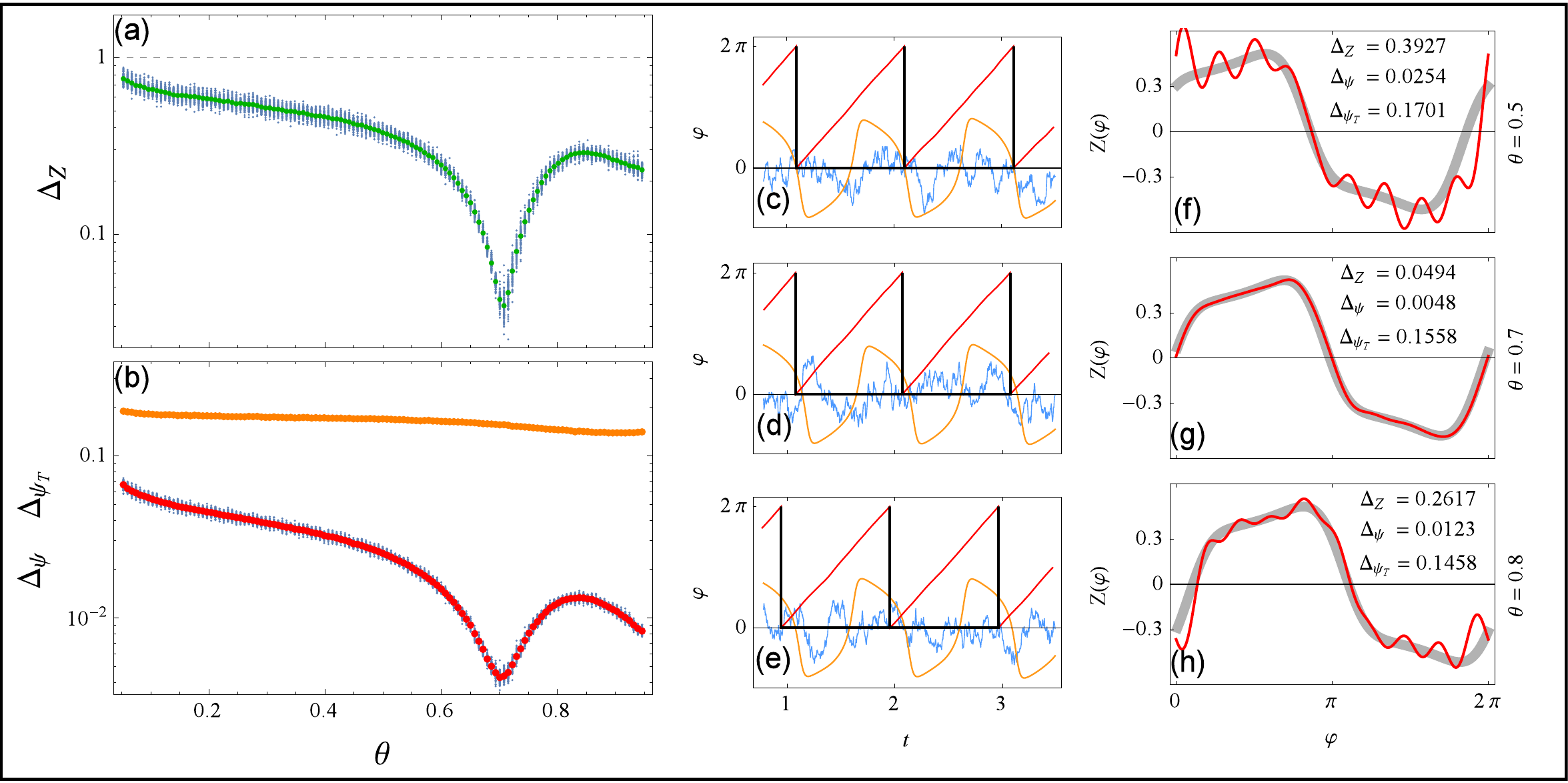}
\caption{The dependence of errors of reconstruction $\Delta_{Z}$ and $\Delta_{\psi}$, Eqs.~(\ref{err_prc}) and (\ref{err_psi}), on the chosen Poincar{\'e} section, determined by 
the threshold parameter $\theta$ (see Methods), 
for the van der Pol oscillator, Eq.~(\ref{vdp}). 
In (a, b) for each value of $\theta$, 100 points corresponding to different realizations of noise are plotted with blue dots and their average in colors: $\Delta_Z$ in green, $\Delta_{\psi}$ in red (and $\Delta_{\psi_T}$ in orange for reference). 
Shown in (c -- e) are segments of signal for three values of $\theta$. Variable $x(t)$ in orange and the input $p(t)$ in blue, scaled for being visually comparable. The times of zero phase $t_m$ are marked with vertical black lines and in red is the reconstructed phase. 
In (f -- h) the corresponding PRC reconstructions are plotted with red. The true PRC is depicted with a thick gray curve. 
Parameters are $\e \lVert Z \rVert = 1$ and $\tau = 0.1$.}
\label{fig:5}
\end{figure}

\begin{figure}[!tbh]
\centering
\includegraphics[width=0.9\columnwidth]{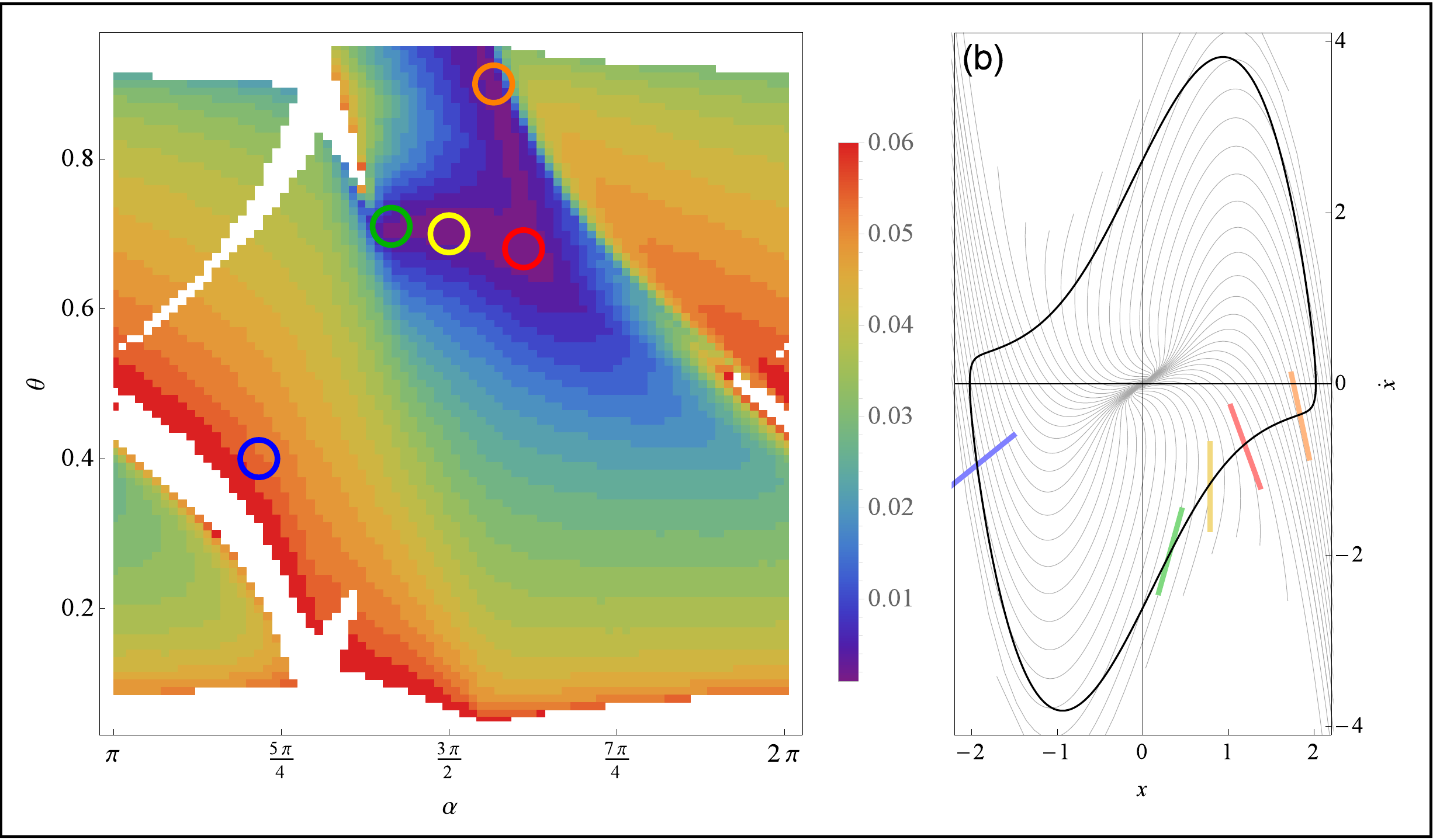}
\caption{The error of reconstruction $\Delta_\psi$, Eq.~(\ref{err_psi}), with respect to the inclination of the Poincar{\'e} surface of section $\alpha$, and its relative position $\theta$, see Methods for definition. 
In (a) the white regions correspond to error values out of range $\Delta_\psi > 0.06$. The particular values marked with colored circles correspond to example surfaces of section plotted in (b) with lines of the same color. 
The red ($\Delta_\psi = 0.0045$), green ($\Delta_\psi = 0.0051$), yellow ($\Delta_\psi = 0.0049$) and orange ($\Delta_\psi = 0.0051$) sections yield good period determination: 
the error is low and in (b) we see that they have a similar inclination to the local isochrones, plotted with thin gray curves. 
The blue one ($\Delta_\psi = 0.054$) is an example of an inaccurate surface of section that does not correspond to local isochrones (see (b)) and therefore yields a high error. 
Notice that the yellow line corresponds to the optimal section in Fig.~\ref{fig:5}.  
The search was performed using a single simulation run of length $t_\text{sim} = 500$ (corresponding to roughly 500 periods). 
Parameters are $\e \lVert Z \rVert = 1$ and $\tau = 0.1$.}
\label{fig:2Dsearch}
\end{figure}

As a first step we probe different threshold values for the signal $x(t)$ 
and choose the one which yields the smallest error $\Delta_\psi$. For this purpose we parametrize 
the threshold values with a parameter $0 < \theta < 1$, see Methods for details, and perform a search
over $\theta$. 
For each value of the threshold $\theta$, 100 simulations with different noise realizations were performed. 
We have calculated both the error of the PRC and the $\Delta_{\psi}$ error. 
We underline, that $\Delta_{\psi}$ can be calculated from data alone, without any \textit{a priori} 
knowledge of the system. The results are shown in Fig.~\ref{fig:5}. 
A clear minimum in both errors around the value  $\theta = 0.7$ indicates the  
optimal threshold that corresponds to the surface of section tangential to the local 
isochrone, see Fig.~\ref{fig:2Dsearch}b.

Generally, the goodness of the surface of section (a line in this example) is determined by two factors:
the line should be tangential to an isochrone and it should be chosen in the domain of fast motion. 
This cannot always be ensured by simple thresholding, $x(t_m)=\text{const}$.
Therefore, we consider inclined lines of section, corresponding to local linear approximation of isochrones.
This can be done even if we have access to only one variable, in our case $x(t)$. 
As is well-known, a phase portrait, topologically equivalent to the true one, can
be obtained from a single time series via computation of the derivative or time-delayed
embedding~\cite{geometry-of-timeseries, embedology}. 
We use here the former option, computing $\dot x(t)$ using the five point finite difference 
scheme~\cite{numerical-derivative}. 
In the two-dimensional embedding with coordinates $(x,\dot x)$ we then search for an optimal linear section,
determined by its position $\theta$ and inclination $\alpha$,  see Methods for details. 
We use a single simulated time series with the total length of 500 periods, mimicking a real world scenario with limited collected data available, 
and compute the error $\Delta_\psi$, for each pair of values $\theta,\alpha$. 
In this way a possibly better surface of section is found, see Fig.~\ref{fig:2Dsearch}.  
This approach can be further improved by a high-dimensional search, e.g., by locally 
approximating isochrones with a higher order polinomial.

\subsubsection*{Comparison with the WSTA technique}
\begin{figure}[!tbh]
\centering
\includegraphics[width=0.65\columnwidth]{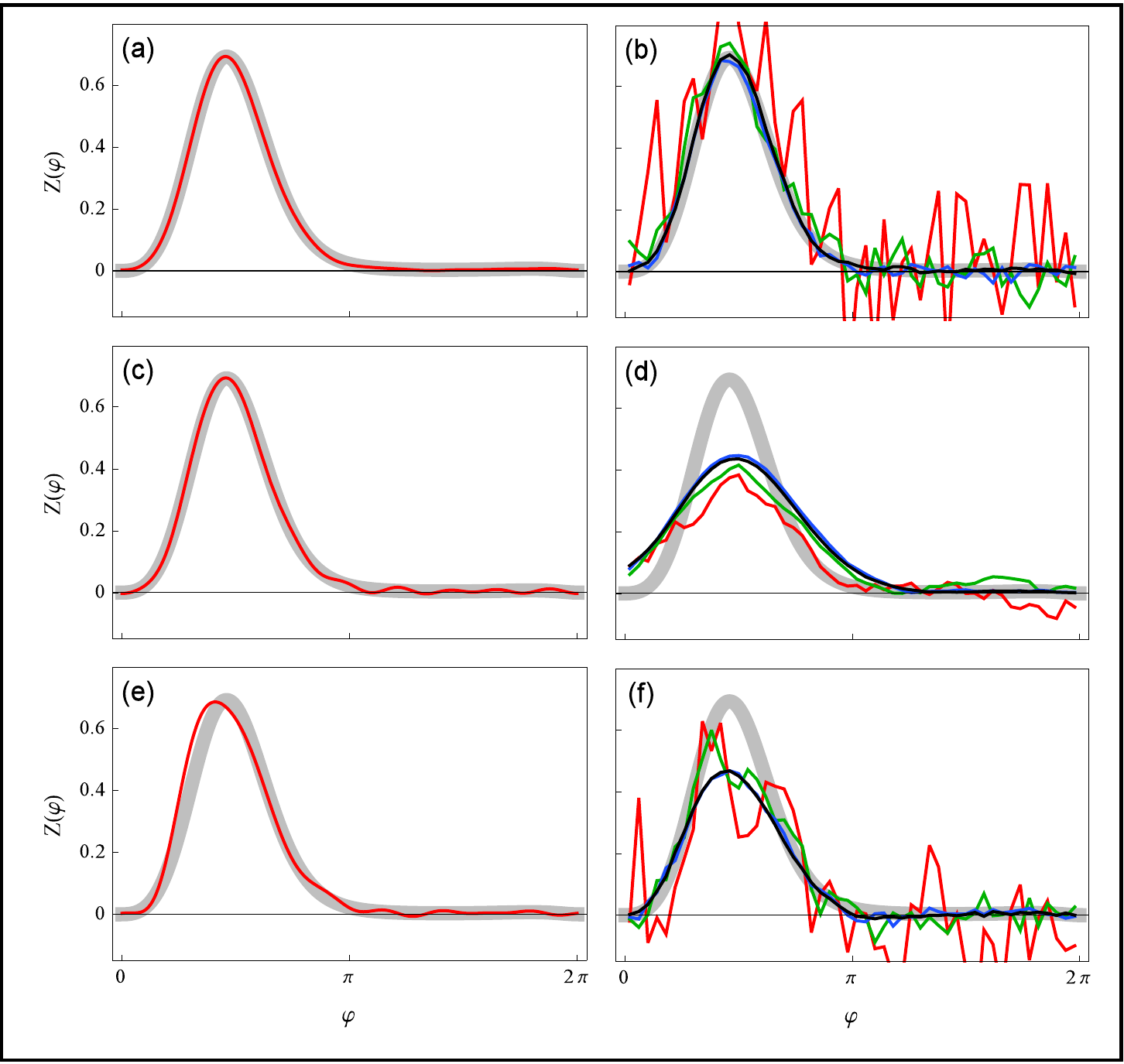}
\caption{Comparison of our method (a, c, e) and the WSTA method~\cite{Ota-PRL-2009, Imai-2017} (b, d, f). 
The data was simulated using the phase model (\ref{eq1}). 
Different colors correspond to the length of time series used for the reconstruction: 
$t_\text{sim} = 10^2$ in red, $10^3$ in green, $10^4$ in blue and $10^5$ in black 
(for our method (a, c, e) only the $t_\text{sim} = 10^2$ is plotted). 
The true PRC is depicted with a thick gray curve. 
In (a, b) the coupling is relatively weak, $\e \lVert Z \rVert = 5$, and the correlation time of the 
input is relatively small, $\tau = 0.01$. These are good conditions for both methods, and indeed both 
perform well. 
In (c, d) the coupling remains as before but the correlation time is larger, $\tau = 0.1$. 
In (e, f) the coupling strength is increased, $\e \lVert Z \rVert = 20$, while the correlation 
time is small again, $\tau = 0.01$. 
}
\label{fig:comparison}
\end{figure}

In the last test we compare the performance of out approach with that of the WSTA 
method~\cite{Ota-PRL-2009,Imai-2017}, see Fig.~\ref{fig:comparison}.
As expected, our technique performs better if the correlation time and/or amplitude of the input is 
relatively large. This is due to the fact that, in contradistinction to WSTA, we do not use the assumptions
of delta-correlated input and of linearly growing phase. Moreover, the test demonstrates that our technique 
works better for shorter time series.

\section*{Discussion}

To summarize,
in this paper we introduce a new method for obtaining the PRC and natural frequency of an oscillator 
relying only on observation of the oscillator's signal and its continuous perturbation. 
We demonstrate its efficiency by recovering the PRC of several model systems driven by correlated noise, 
from only a few hundred observed periods. 
Furthermore, we provide error measures indicating the quality of the inference that can be calculated 
from data, without any additional knowledge of the system. 

We then explore the effects of the amplitude and correlation time of the driving signal. 
Generally, the reconstruction works best for perturbation with low amplitude and short 
correlation time, although the requirements here are not so crucial as in the case of the 
WSTA technique~\cite{Ota-PRL-2009, Imai-2017}. 
The degradation of the model inference at higher amplitudes can be due to two reasons:
the phase description of an oscillator in terms of Eq.~(\ref{eq1}) loses its validity
or/and linear approximation of the phase dynamics becomes too poor for the iterative scheme to converge. 
The latter problem may be solved by using a different initial estimate of the phase, e.g., obtained
via the Hilbert transform.
On the other hand, the degradation of the reconstruction procedure at large correlation times 
occurs due to a limitation of our technique. Indeed, if the driving is very slow and can be approximately
considered as constant within one oscillation period, then the integrals in Eq.~(\ref{eq4}) vanish and 
the whole procedure fails. However, for some realizations of noise, relatively strong and slow driving 
yields a good reconstruction as well 
(see examples in Supplementary material, Fig.~\ref{sup:strong} and Fig.~\ref{sup:slow}) 
and the introduced error measures allow us to examine whether the reconstruction is good or not. 
Finally, we mention that in the white noise limit an additional term, proportional to the square 
of noise intensity, appears in Eq.~(\ref{eq1})~\cite{Teramae-PRL-2009}. This fact can be another source of
error for large perturbation amplitudes. 

Finally, we analyze the importance of proper determination of the points that can be considered as
beginning of the periods, i.e., the points that are assigned zero phase.
This tends to be straightforward for most neuronal models where spikes are the natural choice both 
due to their detection being robust to noise and the fact that during a spike the phase gradient is 
typically small dismissing the importance of the shape of the chosen surface of section. 
In general though, the choice of an appropriate surface of section is crucial. 
To this end we propose to search for an optimal Poincar\'e section which minimizes the reconstruction 
error $\Delta_\psi$. For this search we parameterize these sections by two quantities, namely their 
position and inclination, use a two-dimensional embedding of the measured time series, and vary
these quantities to find the combination that yields minimal $\Delta_\psi$.
In this way we find local approximations of the isochrones by straight lines
(cf Fig.~\ref{sup:iso_estim} in Supplementary material). 
This approach can be further extended to a nonlinear fit of isochrones.
For high-dimensional systems the performance of the PRC estimation can be further improved by 
using a three-dimensional embedding and approximating the isochrones by inclined planes, etc.
Alternatively, for a long data set one can reveal the isochrone structure by estimating 
the surfaces with the mean first return time equal to the period~\cite{Pikovsky-isochrones}.
Next, our approach can be also combined with the technique suggested in the framework of WSTA 
in Ref.~\cite{Imai-2017}. 
There the authors addressed the problem of choosing the proper section by rescaling and averaging 
$n>1$ consecutive periods instead of one. The inaccuracy due to deviation of the used Poincar\'e section
from the true isochrone is then distributed over $n$ periods.
For our approach the inter-event intervals $T_m = t_{m+1}-t_m$ shall be simply replaced by 
${T^{(n)}_m} = t_{m+n}-t_m$, and the left hand side of Eq.~(\ref{eq4}) shall be changed to 
$n\cdot 2\pi$. 
However, other numerical errors accumulate, and therefore we do not expect this approach 
to be superior than the search for an optimal section.

\section*{Methods}

\subsection*{Reconstruction errors}
For quantification of how well our reconstructed PRC resembles the true one we use the following expression
\begin{equation}
	\Delta_Z = \left(\lVert Z^\text{(t)}-Z^\text{(r)}\rVert\right)/\lVert Z^\text{(t)}\rVert\;,  
	\label{err_prc}
\end{equation}
where $Z^{\text{(t)}}$ refers to the true PRC, $Z^{\text{(r)}}$ to the reconstructed one 
and $\lVert \cdot \rVert$ is the $L_2$ function norm: $\lVert f\rVert = \Big[ \int\limits_0^{2\pi}f^2(\varphi)\dd\varphi \Big]^{1/2}$. 
When the reconstructed PRC nearly coincides with the true one the error is small, i.e., $\Delta_Z \ll 1$. 
When the two PRCs are not similar, but are of the same order of magnitude, $\Delta_Z$ is of the order of 1, 
but in general $\Delta_Z$ can have arbitrarily large values. 
For computation of the true PRC we first determine the period of the autonomous limit cycle oscillation,
$T_0$, using the Henon trick~\cite{Henon-82}. 
Next, similarly to \cite{Imai-2017},
we instantaneously perturb the oscillator at different 
phases within one period (phases are taken proportionally to the time from the beginning of the cycle)
and look for the shift in the asymptotic phase. 
If the period where perturbation arrives is denoted by $T_1$ and the following periods 
as $T_2$, $T_3$, $\ldots$, then the phase shift normalized to the perturbation strength $\e$ is 
\begin{equation}
Z^{\text{(t)}}(\vp)=2\pi\frac{nT_0-\sum_{i=1}^n T_i}{\e T_0}\;.
\end{equation}
The number of periods $n$ required for a complete relaxation to the limit cycle can be easily chosen
by trial and error by checking that the result does not depend on $n$.

Next, we quantify how well our reconstruction describes the system which generated the data. 
$\psi_m$ is defined as the reconstructed phase at the end of interval $m$ which in general differs from 
the true phase of $2\pi$ 
due to the inaccuracy of the reconstructed PRC and $\w$. 
Therefore the average deviation of $\psi_m$ from $2\pi$ is a natural measure of goodness for our reconstruction 
\begin{equation}
	\Delta_\psi = \Big\langle \big(\psi_m-2\pi\big)^2 \Big\rangle^{1/2}\;,
	\label{err_psi}
\end{equation}
where $\langle \cdot \rangle$ refers to averaging over $m$. 
It can be computed solely from data (unlike $\Delta_Z$ which requires the knowledge of the true PRC) meaning 
that it can be used in real experiments.
However, $\Delta_\psi$ should not be taken as an absolute measure because it also depends on the inherent 
irregularity of inter-event intervals. 
With that in mind we define a measure of event irregularity 
\begin{equation}
	\Delta_{\psi_T} =  \Big\langle \big( \langle \w \rangle T_m -2\pi \big)^2 \Big\rangle^{1/2}\;,
	\label{err_psi0}
\end{equation}
where $\langle \w \rangle = \langle \frac{2\pi}{T_m} \rangle$ is the mean frequency.
This measure serves as a good benchmark since it 
can be understood as $\Delta_\psi$ calculated for a trivial reconstruction, namely, 
when average period is used as a prediction of next inter-event interval.
Hence, an indication of good reconstruction is the condition $\Delta_\psi \ll \Delta_{\psi_T}$.

\subsection*{A choice of Poincar{\'e} section}
In the simplest case the same phase is assigned to the instants when the signal crosses a 
chosen threshold from above
(the crossings from bellow can be taken as well so long as we are consistent). 
Practically, this instants are determined by linear interpolation between the closest values
above and below the threshold. 
In this work we parameterize a threshold value $s_\text{thr}$ by $\theta \in (0,1)$, so that  
$s_\text{thr} = s_\text{min} + \theta (s_\text{max} - s_\text{min})$,
where $s_\text{min}$ and $s_\text{max}$ are the minimum and maximum of the considered signal. 
With the embedded signal, $x(t)$, $\hat{x}(t)$, we choose a Poincar{\'e} section as a straight 
line at an angle $\alpha$ (in our case we use $\hat{x} = \dot{x}$). This is accomplished by rotating the embedded limit cycle by the angle $\alpha$, such that the 
line of section becomes parallel to the horizontal axis and then taking the 
vertical axis projection of the limit cycle as an auxiliary signal:
$s_{\text{aux}} = -x(t)\sin(\alpha) + \hat{x}(t)\cos(\alpha)$.
Then the auxiliary signal is thresholded as before. 
A depiction of the process may be found in Supplementary material, Fig.~\ref{sup:embedding}.

\newpage
\section*{Author contributions statement}
R.C. and M.R. designed the algorithms and performed the computations. R.C. computed the statistics and prepared 
final figures. Both authors wrote the manuscript.

\section*{Acknowledgements}
We thank Arkady Pikovsky, Igor Goychuk and Chunming Zheng for useful discussions.
The work was supported by ITN COSMOS (funded by
the European Union Horizon 2020 research and innovation
programme under the Marie Sklodowska-Curie grant agreement No. 642563).
Numerical part of this work conducted by M.R. was
supported by the Russian Science Foundation
(Project No. 14-12-00811). We acknowledge the support of the Deutsche Forschungsgemeinschaf and Open Access Publishing Fund of University of Potsdam. 

\bibliographystyle{apsrev4-1}
\bibliography{reconst}

\begin{thebibliography}{10}
\expandafter\ifx\csname url\endcsname\relax
  \def\url#1{\texttt{#1}}\fi
\expandafter\ifx\csname urlprefix\endcsname\relax\def\urlprefix{URL }\fi
\providecommand{\bibinfo}[2]{#2}
\providecommand{\eprint}[2][]{\url{#2}}

\bibitem{Winfree-80}
\bibinfo{author}{Winfree, A.~T.}
\newblock \emph{\bibinfo{title}{The Geometry of Biological Time}}
  (\bibinfo{publisher}{Springer}, \bibinfo{address}{Berlin},
  \bibinfo{year}{1980}).

\bibitem{Glass-Mackey-88}
\bibinfo{author}{Glass, L.} \& \bibinfo{author}{Mackey, M.~C.}
\newblock \emph{\bibinfo{title}{From Clocks to Chaos: {T}he Rhythms of Life.}}
  (\bibinfo{publisher}{Princeton Univ. Press}, \bibinfo{address}{Princeton,
  NJ}, \bibinfo{year}{1988}).

\bibitem{Rinzel-Ermentrout-98}
\bibinfo{author}{Rinzel, J.} \& \bibinfo{author}{Ermentrout, B.}
\newblock \bibinfo{title}{Analysis of neural excitability and oscillations}.
\newblock In \bibinfo{editor}{Koch, C.} \& \bibinfo{editor}{Segev, I.} (eds.)
  \emph{\bibinfo{booktitle}{Methods of Neuronal Modeling}},
  \bibinfo{pages}{251--292} (\bibinfo{publisher}{MIT Press},
  \bibinfo{address}{Cambridge}, \bibinfo{year}{1998}).

\bibitem{PRC-Scholarpedia}
\bibinfo{author}{Canavier, C.~C.}
\newblock \bibinfo{title}{{P}hase response curve}.
\newblock \emph{\bibinfo{journal}{Scholarpedia}} \textbf{\bibinfo{volume}{1}},
  \bibinfo{pages}{1332} (\bibinfo{year}{2006}).

\bibitem{Izhikevich-2007}
\bibinfo{author}{Izhikevich, E.}
\newblock \emph{\bibinfo{title}{The Dynamical Systems in Neuroscience: Geometry
  of Excitability and Bursting}} (\bibinfo{publisher}{The MIT Press},
  \bibinfo{year}{2007}).

\bibitem{Guevara-1986}
\bibinfo{author}{Guevara, M.~R.}, \bibinfo{author}{Shrier, A.} \&
  \bibinfo{author}{Glass, L.}
\newblock \bibinfo{title}{Phase resetting of spontaneously beating embryonic
  ventricular heart cell aggregates}.
\newblock \emph{\bibinfo{journal}{American Journal of Physiology}}
  \textbf{\bibinfo{volume}{251}}, \bibinfo{pages}{1298--1305}
  (\bibinfo{year}{1986}).

\bibitem{Imai-2017}
\bibinfo{author}{Imai, T.}, \bibinfo{author}{Ota, K.} \&
  \bibinfo{author}{Aoyagi, T.}
\newblock \bibinfo{title}{Robust measurements of phase response curves realized
  via multicycle weighted spike-triggered averages}.
\newblock \emph{\bibinfo{journal}{Journal of the Physical Society of Japan}}
  \textbf{\bibinfo{volume}{86}}, \bibinfo{pages}{024009}
  (\bibinfo{year}{2017}).

\bibitem{Cestnik-Rosenblum-17}
\bibinfo{author}{Cestnik, R.} \& \bibinfo{author}{Rosenblum, M.}
\newblock \bibinfo{title}{Reconstructing networks of pulse-coupled oscillators
  from spike trains}.
\newblock \emph{\bibinfo{journal}{Physical Review E}}
  \textbf{\bibinfo{volume}{96}}, \bibinfo{pages}{012209}
  (\bibinfo{year}{2017}).

\bibitem{Galan-PRL_2007}
\bibinfo{author}{Ermentrout, G.~B.}, \bibinfo{author}{Gal\'an, R.~F.} \&
  \bibinfo{author}{Urban, N.~N.}
\newblock \bibinfo{title}{Relating neural dynamics to neural coding}.
\newblock \emph{\bibinfo{journal}{Physical Review Letters}}
  \textbf{\bibinfo{volume}{99}}, \bibinfo{pages}{248103}
  (\bibinfo{year}{2007}).

\bibitem{Ota-PRL-2009}
\bibinfo{author}{Ota, K.}, \bibinfo{author}{Nomura, M.} \&
  \bibinfo{author}{Aoyagi, T.}
\newblock \bibinfo{title}{Weighted spike-triggered average of a fluctuating
  stimulus yielding the phase response curve}.
\newblock \emph{\bibinfo{journal}{Physical Review Letters}}
  \textbf{\bibinfo{volume}{103}}, \bibinfo{pages}{024101}
  (\bibinfo{year}{2009}).

\bibitem{Morinaga-2015}
\bibinfo{author}{Morinaga, K.}, \bibinfo{author}{Miyata, R.} \&
  \bibinfo{author}{Aonishi, T.}
\newblock \bibinfo{title}{Optimal colored noise for estimating phase response
  curves}.
\newblock \emph{\bibinfo{journal}{Journal of the Physical Society of Japan}}
  \textbf{\bibinfo{volume}{84}}, \bibinfo{pages}{094801}
  (\bibinfo{year}{2015}).

\bibitem{Gabor-Hilbert}
\bibinfo{author}{Gabor, D.}
\newblock \bibinfo{title}{Theory of communication}.
\newblock \emph{\bibinfo{journal}{Journal of Institution of Electrical
  Engineers}} \textbf{\bibinfo{volume}{93}}, \bibinfo{pages}{429--457}
  (\bibinfo{year}{1946}).

\bibitem{Hilbert-Scholarpedia}
\bibinfo{author}{Freeman, W.~J.}
\newblock \bibinfo{title}{{H}ilbert transform for brain waves}.
\newblock \emph{\bibinfo{journal}{Scholarpedia}} \textbf{\bibinfo{volume}{2}},
  \bibinfo{pages}{1338} (\bibinfo{year}{2007}).

\bibitem{Pikovsky-Rosenblum-Kurths-01}
\bibinfo{author}{Pikovsky, A.}, \bibinfo{author}{Rosenblum, M.} \&
  \bibinfo{author}{Kurths, J.}
\newblock \emph{\bibinfo{title}{Synchronization. A Universal Concept in
  Nonlinear Sciences.}} (\bibinfo{publisher}{Cambridge University Press},
  \bibinfo{address}{Cambridge}, \bibinfo{year}{2001}).

\bibitem{Kralemann_et_al-08}
\bibinfo{author}{Kralemann, B.}, \bibinfo{author}{Cimponeriu, L.},
  \bibinfo{author}{Rosenblum, M.}, \bibinfo{author}{Pikovsky, A.} \&
  \bibinfo{author}{Mrowka, R.}
\newblock \bibinfo{title}{Phase dynamics of coupled oscillators reconstructed
  from data}.
\newblock \emph{\bibinfo{journal}{Phys. Rev. E}} \textbf{\bibinfo{volume}{77}},
  \bibinfo{pages}{066205} (\bibinfo{year}{2008}).

\bibitem{Galan-Ermentrout-Urban-05}
\bibinfo{author}{Gal\'an, R.~F.}, \bibinfo{author}{Ermentrout, G.~B.} \&
  \bibinfo{author}{Urban, N.~N.}
\newblock \bibinfo{title}{Efficient estimation of phase-resetting curves in
  real neurons and its significance for neural-network modeling}.
\newblock \emph{\bibinfo{journal}{Phys. Rev. Lett.}}
  \textbf{\bibinfo{volume}{94}}, \bibinfo{pages}{158101}
  (\bibinfo{year}{2005}).

\bibitem{Kralemann_et_al-13}
\bibinfo{author}{Kralemann, B.} \emph{et~al.}
\newblock \bibinfo{title}{In vivo cardiac phase response curve elucidates human
  respiratory heart rate variability}.
\newblock \emph{\bibinfo{journal}{Nature Communications}}
  \textbf{\bibinfo{volume}{4}}, \bibinfo{pages}{2418} (\bibinfo{year}{2013}).

\bibitem{Hansel-95}
\bibinfo{author}{Hansel, D.}, \bibinfo{author}{Mato, G.} \&
  \bibinfo{author}{Meunier, C.}
\newblock \bibinfo{title}{Synchrony in excitatory neural networks}.
\newblock \emph{\bibinfo{journal}{Neural Computation}}
  \textbf{\bibinfo{volume}{7}}, \bibinfo{pages}{307--337}
  (\bibinfo{year}{1995}).

\bibitem{ornstein-uhlenbeck}
\bibinfo{author}{Uhlenbeck, G.~E.} \& \bibinfo{author}{Ornstein, L.~S.}
\newblock \bibinfo{title}{On the theory of {B}rownian motion}.
\newblock \emph{\bibinfo{journal}{Physical review}}
  \textbf{\bibinfo{volume}{36}}, \bibinfo{pages}{823–841}
  (\bibinfo{year}{1930}).

\bibitem{Morris-Lecar-81}
\bibinfo{author}{Morris, C.} \& \bibinfo{author}{Lecar, H.}
\newblock \bibinfo{title}{Voltage oscillations in the barnacle giant muscle
  fiber}.
\newblock \emph{\bibinfo{journal}{Biophys. J.}} \textbf{\bibinfo{volume}{35}},
  \bibinfo{pages}{193 -- 213} (\bibinfo{year}{1981}).

\bibitem{Guckenheimer-75}
\bibinfo{author}{Guckenheimer, J.}
\newblock \bibinfo{title}{Isochrons and phaseless sets}.
\newblock \emph{\bibinfo{journal}{Journal of Mathematical Biology}}
  \textbf{\bibinfo{volume}{1}}, \bibinfo{pages}{259–273}
  (\bibinfo{year}{1975}).

\bibitem{van-der-Pol-26}
\bibinfo{author}{van~der Pol, B.}
\newblock \bibinfo{title}{On relaxation-oscillations}.
\newblock \emph{\bibinfo{journal}{Philosophical Magazine}}
  \textbf{\bibinfo{volume}{2}}, \bibinfo{pages}{978--992}
  (\bibinfo{year}{1926}).

\bibitem{geometry-of-timeseries}
\bibinfo{author}{Packard, N.~H.}, \bibinfo{author}{Crutchfield, J.~P.},
  \bibinfo{author}{Farmer, J.~D.} \& \bibinfo{author}{Shaw, R.~S.}
\newblock \bibinfo{title}{Geometry from a time series}.
\newblock \emph{\bibinfo{journal}{Physical Review Letters}}
  \textbf{\bibinfo{volume}{45}}, \bibinfo{pages}{712--716}
  (\bibinfo{year}{1980}).

\bibitem{embedology}
\bibinfo{author}{Sauer, T.}, \bibinfo{author}{Yorke, J.~A.} \&
  \bibinfo{author}{Casdagli, M.}
\newblock \bibinfo{title}{Embedology}.
\newblock \emph{\bibinfo{journal}{Journal of Statistical Physics}}
  \textbf{\bibinfo{volume}{65}}, \bibinfo{pages}{579--616}
  (\bibinfo{year}{1991}).

\bibitem{numerical-derivative}
\bibinfo{author}{Fornberg, B.}
\newblock \bibinfo{title}{Generation of finite difference formulas on
  arbitrarily spaced grids}.
\newblock \emph{\bibinfo{journal}{Mathematics of Computation}}
  \textbf{\bibinfo{volume}{51}}, \bibinfo{pages}{699--706}
  (\bibinfo{year}{1988}).

\bibitem{Teramae-PRL-2009}
\bibinfo{author}{Teramae, J.}, \bibinfo{author}{Nakao, H.} \&
  \bibinfo{author}{Ermentrout, G.~B.}
\newblock \bibinfo{title}{Stochastic phase reduction for a general class of
  noisy limit cycle oscillators}.
\newblock \emph{\bibinfo{journal}{Physical Review Letters}}
  \textbf{\bibinfo{volume}{102}}, \bibinfo{pages}{194102}
  (\bibinfo{year}{2009}).

\bibitem{Pikovsky-isochrones}
\bibinfo{author}{Schwabedal, J. T.~C.} \& \bibinfo{author}{Pikovsky, A.}
\newblock \bibinfo{title}{Phase description of stochastic oscillations}.
\newblock \emph{\bibinfo{journal}{Physical Review Letters}}
  \textbf{\bibinfo{volume}{110}}, \bibinfo{pages}{204102}
  (\bibinfo{year}{2013}).

\bibitem{Henon-82}
\bibinfo{author}{Henon, M.}
\newblock \bibinfo{title}{On the numerical computation of {P}oincar{\'e} maps}.
\newblock \emph{\bibinfo{journal}{Physica D: Nonlinear Phenomena}}
  \textbf{\bibinfo{volume}{5}}, \bibinfo{pages}{412--414}
  (\bibinfo{year}{1982}).

\bibitem{milstein}
\bibinfo{author}{Milstein, G.~N.}
\newblock \bibinfo{title}{Approximate integration of stochastic differential
  equations}.
\newblock \emph{\bibinfo{journal}{Teoriya Veroyatnostei i ee Primeneniya}}
  \textbf{\bibinfo{volume}{19}}, \bibinfo{pages}{583--588}
  (\bibinfo{year}{1974}).

\bibitem{hindmarsh-rose}
\bibinfo{author}{Hindmarsh, J.~L.} \& \bibinfo{author}{Rose, R.~M.}
\newblock \bibinfo{title}{A model of neuronal bursting using three coupled
  first order differential equations}.
\newblock \emph{\bibinfo{journal}{Proceedings of the Royal Society of London B:
  Biological Sciences}} \textbf{\bibinfo{volume}{221}},
  \bibinfo{pages}{87--102} (\bibinfo{year}{1984}).

\bibitem{isochrone-scholarpedia}
\bibinfo{author}{Josic, K.}, \bibinfo{author}{Shea-Brown, E.~T.} \&
  \bibinfo{author}{Moehlis, J.}
\newblock \bibinfo{title}{{I}sochron}.
\newblock \emph{\bibinfo{journal}{Scholarpedia}} \textbf{\bibinfo{volume}{1}},
  \bibinfo{pages}{1361} (\bibinfo{year}{2006}).

\end{thebibliography}

\vspace*{50px}

\beginsupplement

\section*{Supplementary material}

\subsection*{White noise driving}
\begin{figure}[!htb]
\centering
\includegraphics[width=0.85\columnwidth]{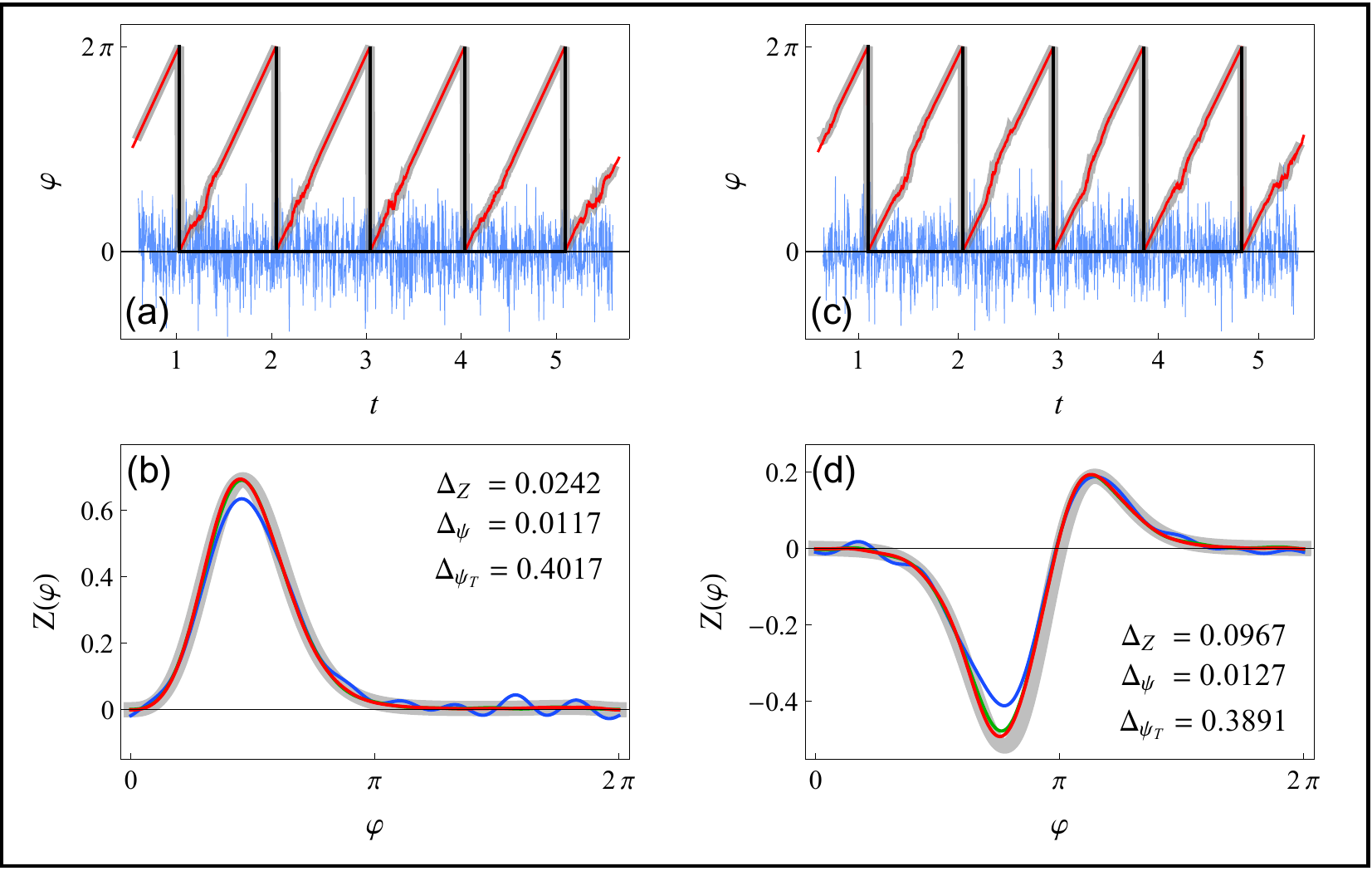}
\caption{A reconstruction for a phase oscillator driven by white noise (analogous to Fig.~\ref{fig:1} in the main text, (a, c) in blue differences of the Wiener process $\Delta W$ instead of signal $p$). The driving strength is $\e \lVert Z \rVert = 1$.}
\label{sup:white}
\end{figure}

Here we demonstrate a reconstruction with Gaussian white noise input $\langle p(t) p(t') \rangle = \e^2 \delta(t-t')$, see Fig.~\ref{sup:white}. 
We simulate the system by integrating Eq.~(\ref{eq1}) in the main text with the Milstein scheme~\cite{milstein} using the PRCs introduced in the main text, Eqs.~(\ref{prc1}) and (\ref{prc2}). 
We record the differences of the Wiener process $\Delta W$ as the driving signal. 

\newpage
\subsection*{The Morris-Lecar neuron: examples of strong: $\e \lVert Z \rVert = 20$, and slow: $\tau = 10$ driving}
\begin{figure}[!htb]
\centering
\includegraphics[width=0.39\columnwidth]{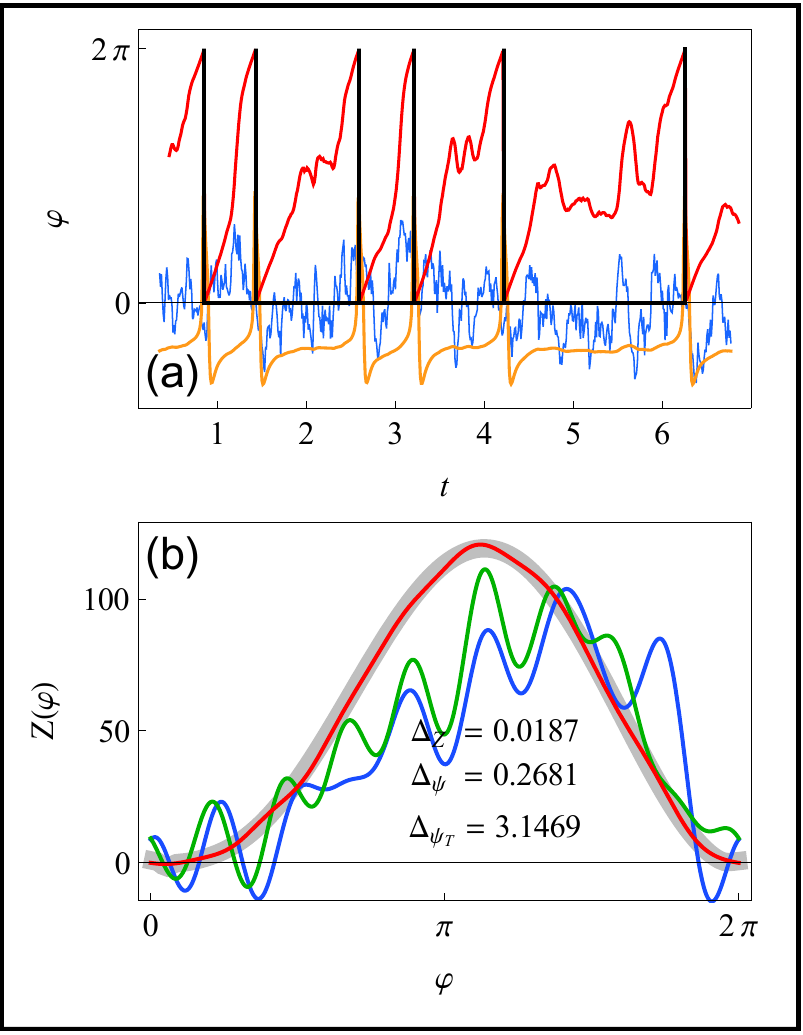}
\caption{An example of a reconstruction with strong driving, Morris-Lecar neuron (similar to Fig.~\ref{fig:1} in the main text). Parameters are $\e \lVert Z \rVert = 20$ and $\tau = 0.1$.}
\label{sup:strong}
\end{figure}

\begin{figure}[!htb]
\centering
\includegraphics[width=0.39\columnwidth]{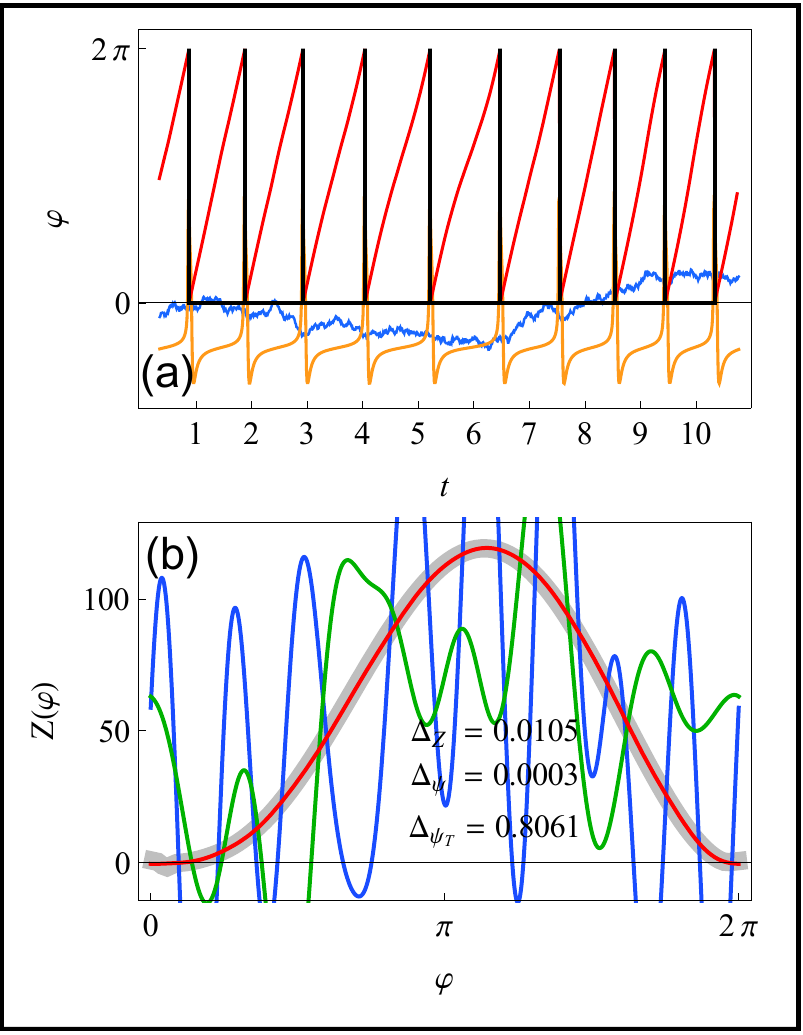}
\caption{An example of a reconstruction with slow driving, Morris-Lecar neuron (similar to Fig.~\ref{fig:1} in the main text). Parameters are $\e \lVert Z \rVert = 3$ and $\tau = 10$.}
\label{sup:slow}
\end{figure}

On average, cases with strong and slow driving yield bad reconstructions, 
however, this can vary from case to case. 
Here we show examples of good reconstructions for those circumstances, see Fig.~\ref{sup:strong} for strong and Fig.~\ref{sup:slow} for slow driving. 
They were chosen as the best from 20 trials (in terms of error measures, Eqs.~(\ref{err_prc}), (\ref{err_psi}) and (\ref{err_psi0}) introduced in the main text). 

\newpage
\subsection*{The Hindmarsh-Rose neuron: example on bursting}
\begin{figure}[!htb]
\centering
\includegraphics[width=0.4\columnwidth]{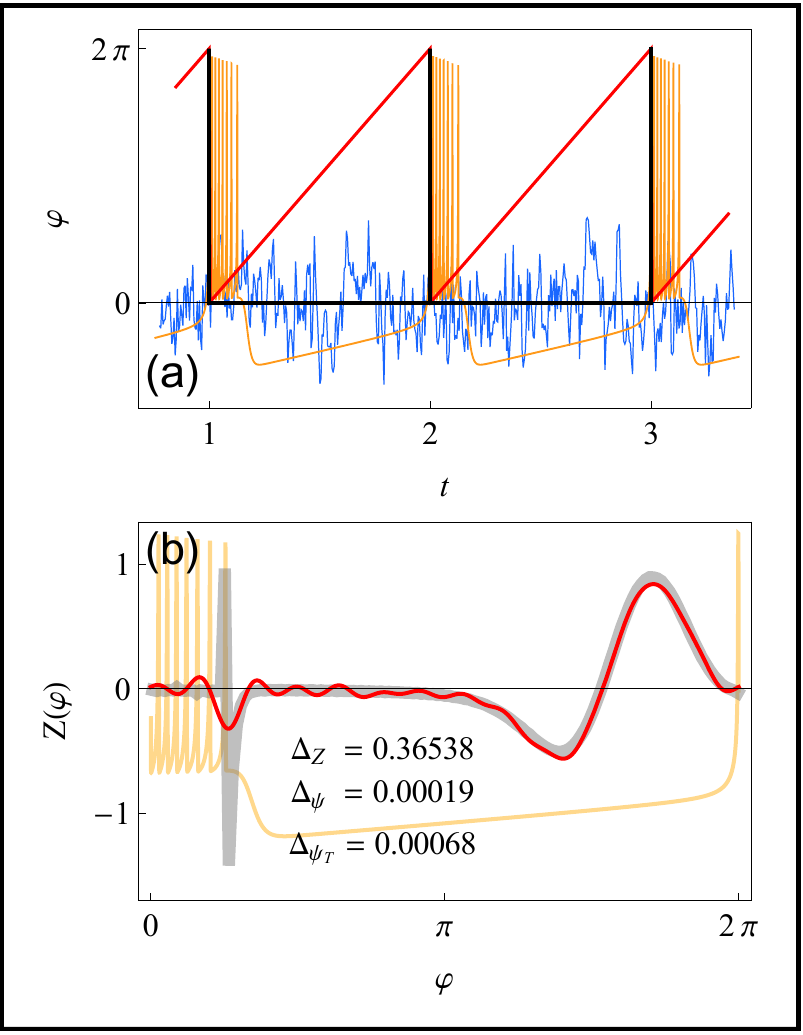}
\caption{An example of a reconstruction using bursts as the events with equal phase, Hindmarsh-Rose neuron (similar to Fig.~\ref{fig:1} in the main text). Parameters are $\e \lVert Z \rVert = 0.01$ and $\tau = 0.01$.}
\label{sup:burst}
\end{figure}

To further support the argument that the data our method requires is general we perform a reconstruction on a bursting neuron. 
We use the Hindmarsh-Rose neuronal model~\cite{hindmarsh-rose}: 
\begin{equation}
\begin{split}
\dot x =& \ I + y -ax^3 +bx^2 - z + p(t)\;, \\
\dot y =& \ c -dx^2 - y\;, \\
\dot z =& \ r[s(x-x_R) - z]\;, 
\end{split}
\label{hr}
\end{equation}
the parameters are:
$I=1.28$,
$a=1$, $b=3$, $c=1$, $d=5$, $r=0.0006$, $s=4$ and $x_R = -1.6$.
In this regime the neuron is bursting periodically. 
For the reconstruction we use parameters $\e \lVert Z \rVert = 0.01$, $\tau = 0.01$, $t_\text{sim} = 1000$, $\dd t = 0.0001$ and $N = 15$. 
The beginning of the period is determined by the time of the first spike in a burst (this is arbitrary, any of the spikes could be considered for this role). 
See Fig.~\ref{sup:burst} for a reconstruction depiction. 
The PRC of this bursting cycle has a very sharp feature around the phase $\varphi \approx 1$ where the bursting stops. 
This is hard to capture with our method since we take the Fourier representation of the PRC and steep features require several harmonics to be expressed. 
Taking many harmonics can be impractical since each harmonic yields an unknown constant needed to be optimized, which can make the data requirements considerably big. 
With $N = 15$ Fourier harmonics that feature is to some degree smeared in phase, but still apparent. 

\newpage
\subsection*{Choice of a proper Poincar{\'e} section}
\begin{figure}[!htpb]
\centering
\includegraphics[width=0.74\columnwidth]{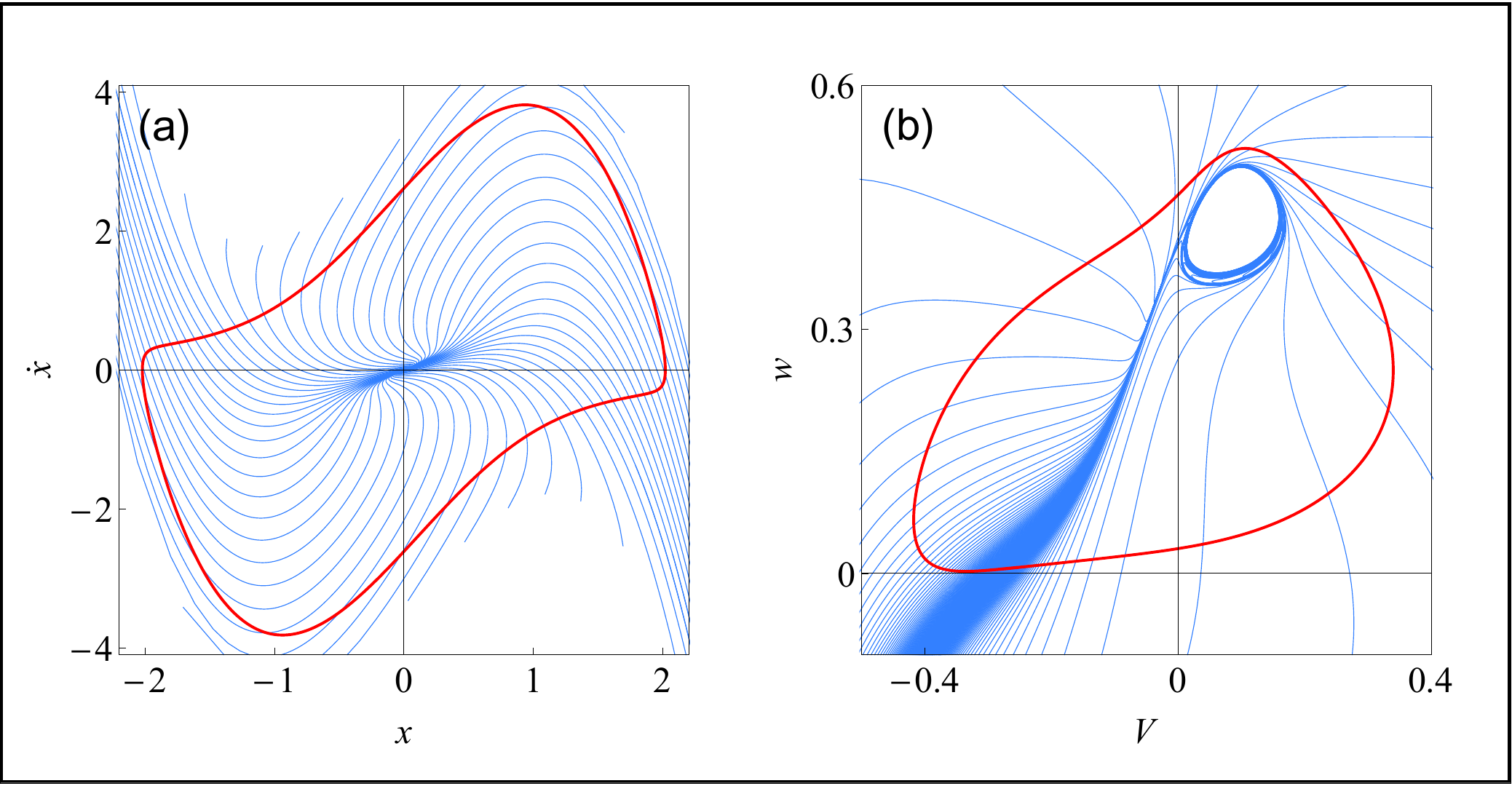}
\caption{The isochrone structure of the two oscillators used in the main text, (a): van der Pol, Eq.~(\ref{vdp}) and (b): Morris-Lecar, Eq.~(\ref{ml}). The limit cycle is plotted with a thick red line while the isochrones with thin blue ones. 
In (a) there are 50 isochrones corresponding to equal phase intervals, while in (b) there are 200.}
\label{sup:isos}
\end{figure}

In the case of a forced oscillator, 
the correct Poincar{\'e} section for determining instants of same phase corresponds to an isochrones surface. 
Here we show the isochrone structure of the two oscillators used in the main text, van der Pol and Morris-Lecar, see Fig.~\ref{sup:isos}. 
The isochrones were computed using the backward integration method~\cite{isochrone-scholarpedia}. 

\begin{figure}[!htpb]
\centering
\includegraphics[width=0.53\columnwidth]{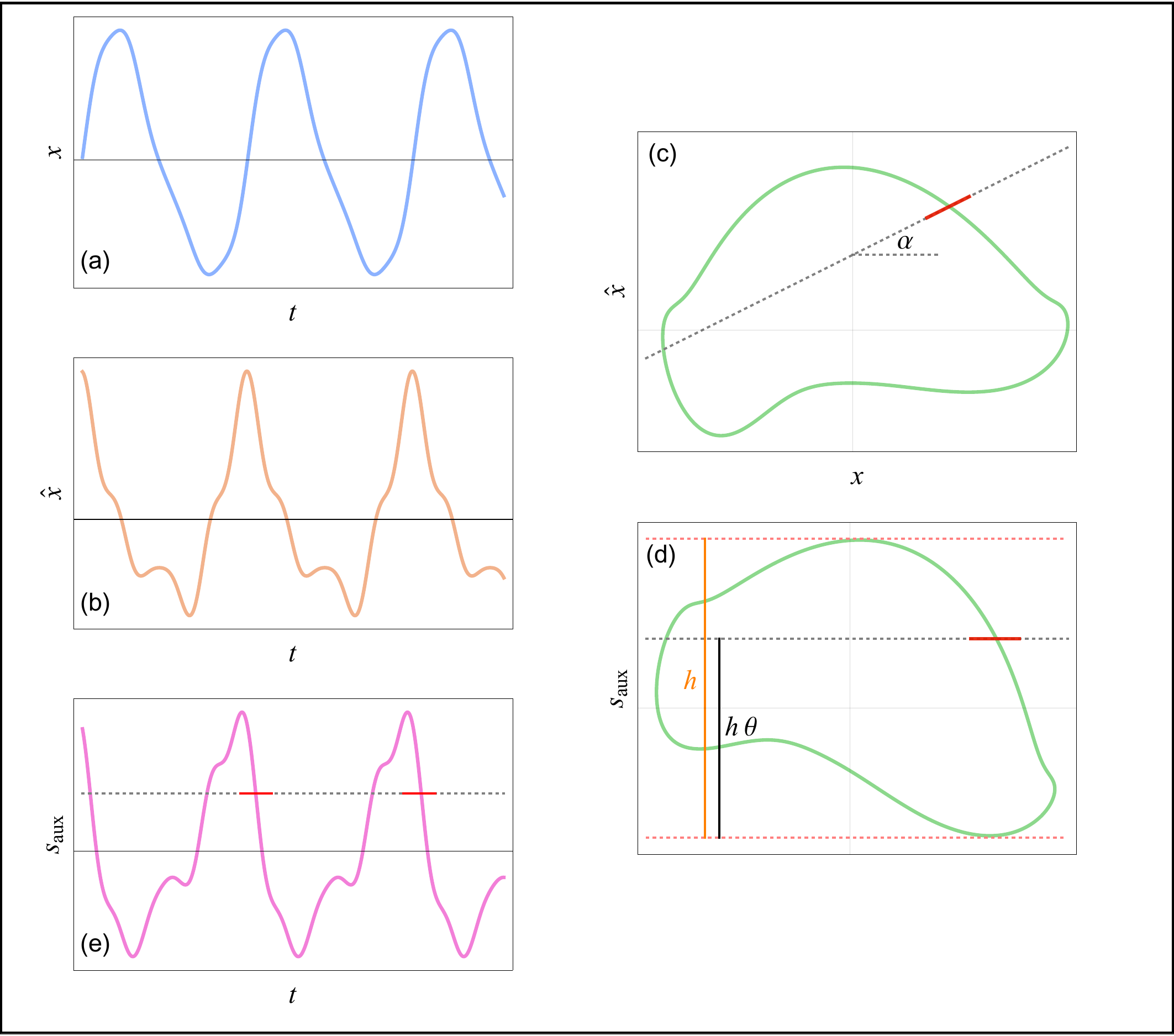}
\caption{A schematic depiction of period determination with the use of signal embedding and inclined surfaces of section, as used for Fig.~\ref{fig:2Dsearch} in the main text. 
In (a) the measured signal $x(t)$ and in (b) the proxi variable $\hat{x}(t)$ (in our case we used the first derivative). 
In (c) the embedded signal (with green) as well as the chosen surface of section depicted with a straight red line. In (d) the embedded signal rotated by $\alpha$ so that the chosen surface of section becomes paralel to the horizontal axis. 
In (e) the vertical projection of the rotated embedded signal which is to be thresholded.}
\label{sup:embedding}
\end{figure}

\noindent
Now suppose we only measure one variable $x(t)$ but want to embed our signal in 2 dimensions and then determine the periods using a Poincar{\'e} section at an arbitrary angle $\alpha$. 
This is done by rotating the embedded limit cycle and calculating the threshold value accordingly, a depiction can be seen in Fig.~\ref{sup:embedding}. 

\newpage
\subsection*{}
\begin{figure}[!htpb]
\centering
\includegraphics[width=0.52\columnwidth]{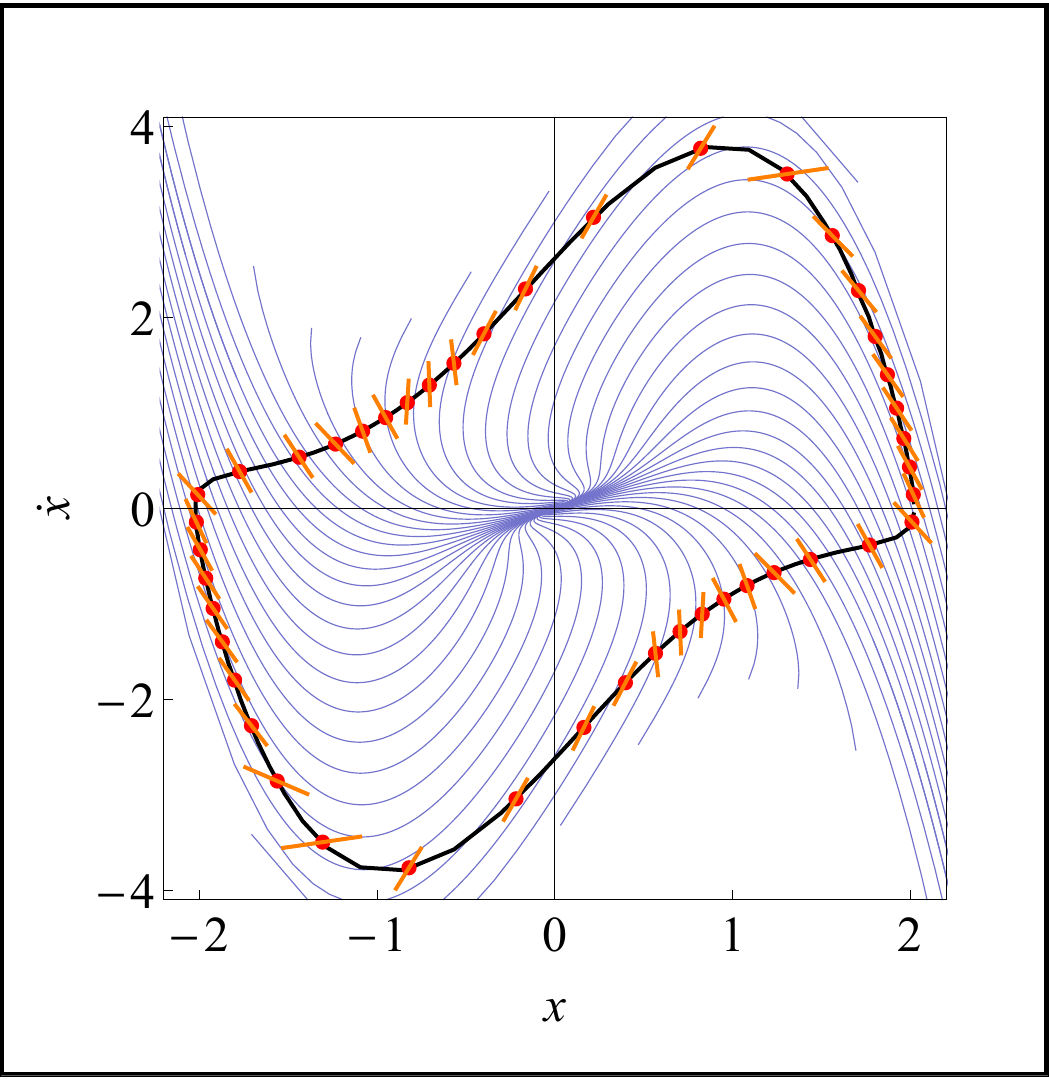}
\caption{An estimated local isochrone structure using only one time series of length $t_\text{sim} = 500$ (coresponding to roughly 500 periods). 
The true isochrones are depicted with thin purple curves. The points where the inclination of the isochrones is estimated (in red) are uniformly distributed in angle. The estimated isochrones are depicted with orange lines. 
Parameters are $\e \lVert Z \rVert = 1$ and $\tau = 0.01$.}
\label{sup:iso_estim}
\end{figure}

Using a similar search to that performed in the main text in Fig.~\ref{fig:2Dsearch}, we can estimate the isochrone structure in the vicinity of the limit cycle, see Fig.~\ref{sup:iso_estim}. 
First, we embed our signl $x(t)$ in two dimensions (see Fig.~\ref{sup:embedding} above). 
Then, considering the embedded signal in polar cooridnates, 
we average the radius variable over the angle variable with finite binning to obtain an approximation of the unpertubed limit cycle. 
On the obtained limit cycle approximation we choose points uniformly distributed in angle and for each point try several Poincar{\'e} sections at different angles. 
The angle corresponding to the lowest error should closely match the local isochrone. 

\begin{figure}[!htb]
\centering
\includegraphics[width=0.9\columnwidth]{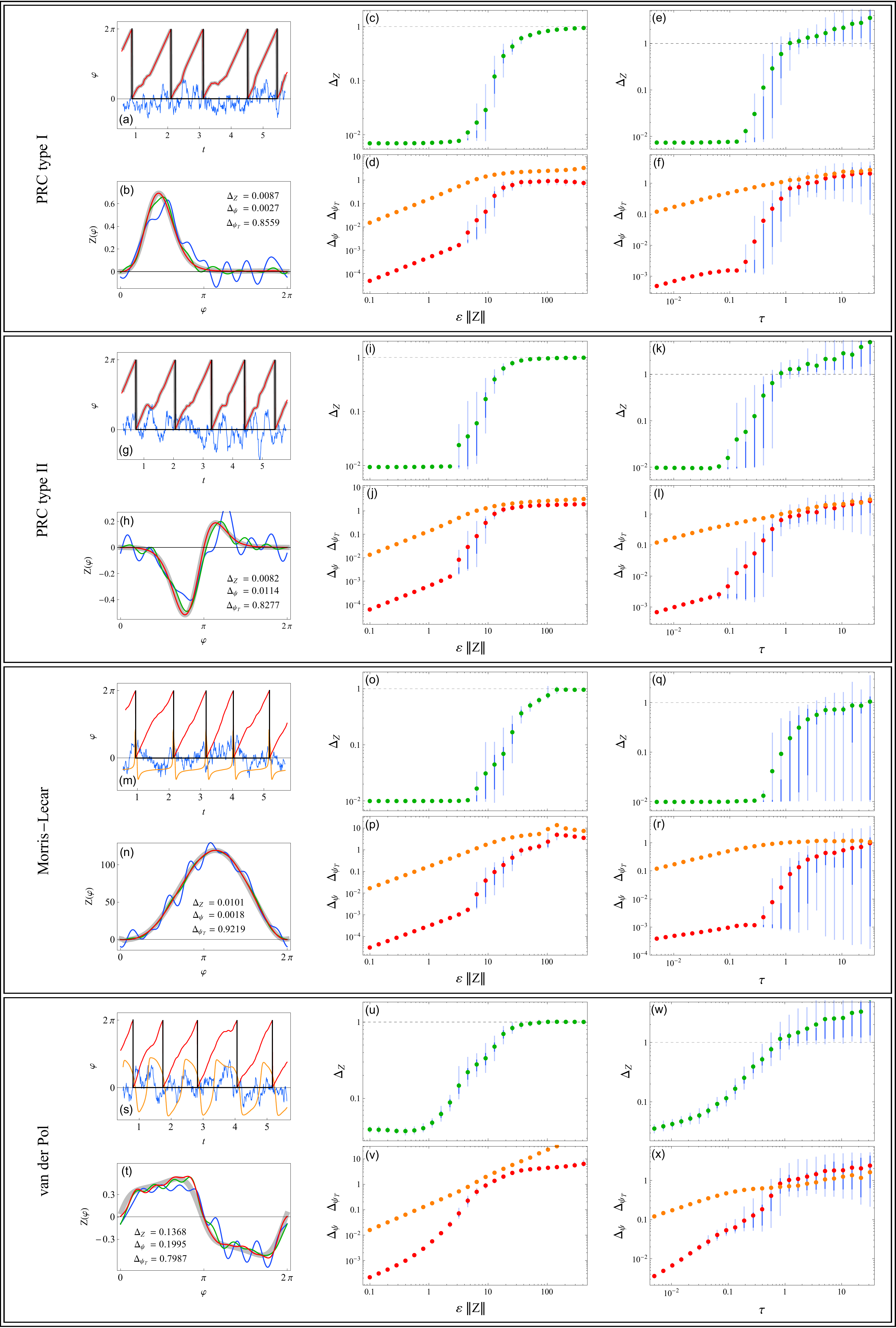}
\caption{Overview comparison of test oscillators: PRC type I, PRC type II, Morris-Lecar and van der Pol. In plots (a, g, m, s) the signal, in (b, h, n, t) the PRC reconstruction, in (c, d, i, j, o, p, u, v) the effect of driving amplitude $\e$ and in (e, f, k, l, q, r, w, x) the effect of driving correlation time $\tau$. 
The parameters are $\e \lVert Z \rVert = 5$ and $\tau = 0.1$ in (a, b, g, h, m, n, s, t), $\tau = 0.1$ in (c, d, i, j, o, p, u, v) and $\e \lVert Z \rVert = 3$ in (e, f, k, l, q, r, w, x).}
\label{sup:all}
\end{figure}

\newpage 
\thispagestyle{empty}

\end{document}